\documentclass[letterpaper]{article} 
\usepackage[]{aaai25}  
\usepackage{amsfonts}
\usepackage{times}  
\usepackage{helvet}  
\usepackage{courier}  
\usepackage[hyphens]{url}  
\usepackage{graphicx} 
\usepackage{natbib}  

\usepackage{xcolor}
\newcommand{\answerYes}[1]{\textcolor{blue}{#1}} 
 
\newcommand{\answerNA}[1]{\textcolor{gray}{#1}}

\usepackage{caption} 
\usepackage{algorithm}
\usepackage{algorithmic}

\usepackage{newfloat}
\usepackage{listings}
\DeclareCaptionStyle{ruled}{labelfont=normalfont,labelsep=colon,strut=off} 
\floatstyle{ruled}
\newfloat{listing}{tb}{lst}{}
\floatname{listing}{Listing}
\usepackage{pgfplots}
\usepgfplotslibrary{fillbetween} 
\usepgfplotslibrary{groupplots}
\usepackage{caption}
\usepackage{subcaption}
\usepackage{pgfplotstable}
\usepackage{xcolor}
\usepackage[abs]{overpic}
\usepackage{multirow}
\usetikzlibrary{intersections}
\usetikzlibrary{patterns}
\usepackage{booktabs}
\usepackage{amsmath}
\usepackage{soul}

\title{Birds of a Feather Cluster Together:\\A Platform-Agnostic Framework for Social Graph Construction}
\title{Data-Efficient, Platform-Agnostic Social Graph Construction\\ for Cross-Platform Analysis: Evidence from the 2024 U.S. Election}
\title{Data-Efficient Platform-Agnostic Social Graph Construction: A Case Study of the 2024 U.S. Election}
\title{Less Data, More Signal: Platform-Agnostic Social Graphs for \\Cross-Platform Analysis of the 2024 U.S. Election}
\title{Tracing Cross-Platform Information Flow in the 2024 U.S. Election:\\A Platform-Agnostic Framework for Social Graph Construction}

\title{Bridging the Narrative Divide:\\ Platform-Agnostic Discourse Graphs Reveal Bridge Users in the 2024 U.S. Election}

\title{Bridging the Narrative Divide:\\ Cross-Platform Discourse Networks and the 2024 U.S. Election}

\title{Bridging the Narrative Divide:\\Cross-Platform Discourse Networks in Fragmented Ecosystems}

\author{
    Patrick Gerard\textsuperscript{\rm 1}, 
    Hans W. A. Hanley\textsuperscript{\rm 2},
    Luca Luceri\textsuperscript{\rm 1},
    Emilio Ferrara\textsuperscript{\rm 3}
}
\affiliations{
    \textsuperscript{\rm 1}Information Sciences Institute, University of Southern California\\
    \textsuperscript{\rm 2}Stanford University\\
    \textsuperscript{\rm 3}Thomas Lord Department of Computer Science, University of Southern California\\
    pgerard@isi.edu, hhanley@cs.stanford.edu,  lluceri@isi.edu,
    emiliofe@usc.edu
}

\usepackage{bibentry}
\usetikzlibrary{pgfplots.statistics}
\usepackage{makecell}

\pgfplotsset{compat=newest} 

\begin{document}

\maketitle

\begin{abstract}
Political discourse has grown increasingly fragmented across different social platforms,  making it challenging to trace how narratives spread and evolve within such a fragmented information ecosystem. Reconstructing social graphs and information diffusion networks is challenging, and available strategies typically depend on platform-specific features and behavioral signals which are often incompatible across systems and increasingly restricted. To address these challenges, we present a platform-agnostic framework that allows to accurately and efficiently reconstruct the underlying social graph of users' cross-platform interactions, based on discovering latent narratives and users' participation therein. 
Our method achieves state-of-the-art performance in key network--based tasks: \textit{ information operation detection}, \textit{ideological stance prediction}, and \textit{cross-platform engagement prediction}---while requiring significantly less data than existing alternatives and capturing a broader set of users. When applied to cross-platform information dynamics between Truth Social and X (formerly Twitter), our framework reveals a small, mixed-platform group of \textit{bridge users}, comprising just 0.33\% of users and 2.14\% of posts, who introduce nearly 70\% of \textit{migrating narratives} to the receiving platform. These findings offer a structural lens for anticipating how narratives traverse fragmented information ecosystems, with implications for cross-platform governance, content moderation, and policy interventions.

\end{abstract}

\begin{links}
    \link{Code}{https://tinyurl.com/CANE2025}
    \link{Data}{https://tinyurl.com/CANE2025}
\end{links}

\section{Introduction}

Political communication today unfolds across a fragmented digital landscape, shaped by the dynamics of ideologically distinct platforms~\cite{zhang2025trump}. During the January 6th, 2021, Capitol insurrection, participants engaged in parallel but intersecting conversations across Parler, X (formerly Twitter), and Gab~\cite{sipka2022comparing,vishnuprasad2024tracking}, which were later amplified on mainstream platforms via news reporting~\cite{zulli2023news,luceri2021social}. More recently, the emergence of Truth Social, a platform created by U.S. President Donald Trump, has introduced a new locus of politically salient discourse~\cite{zhang2025trump, gerard2023truth}, with narratives frequently originating there before diffusing outward~\cite{shah2024unfiltered}. This fragmentation became particularly salient during the 2024 U.S. presidential election, when narratives, unverified claims, and emotionally charged rhetoric moved rapidly across ecosystem boundaries \cite{minici2024uncovering}. Yet despite growing concern, we still lack systematic tools for tracing how discourse migrates between these siloed communities, or for identifying the structures that enable it.

Most traditional network approaches rely on platform-specific behaviors such as reposts, mentions, or follower ties~\cite{cinelli2021echo}, which are often incompatible across ecosystems and increasingly inaccessible due to API restrictions~\cite{tromble2021have}. More recent content-based methods infer user relationships through linguistic similarity~\cite{ng2022cross, luceri2024unmasking}, but typically depend on pairwise message comparisons or token-level co-occurrence~\cite{cinus2024exposing}. As we show in this work, such approaches fail to capture persistent alignment, generate unstable user communities, and generalize poorly across platforms. These limitations obscure the structural architecture of discourse, particularly the aspects that allow narratives to move between fragmented online communities.

To overcome these limitations and understand cross-platform dynamics during the 2024 U.S. Presidential election, we model users not through interaction patterns or surface-level text similarity, but as distributions over latent narratives. Our discourse-centered framework links users through shared participation in semantic clusters of ideas, capturing persistent alignment even without direct ties. This approach disentangles user similarity from platform-specific behavior, allowing for robust and scalable network construction across sparse and fragmented ecosystems.

Empirically, we find that this method offers several advantages. We test its effectiveness across both standard user representation tasks and a new benchmark for cross-platform engagement prediction. In both intra- and inter-platform settings, our method matches or exceeds the performance of existing approaches while requiring significantly less data. It captures a broader set of users, identifies stable communities across platforms, and supports large-scale modeling of discourse. Our method, without relying on platform-specific metadata, not only improves generalization and diffusion modeling across siloed ecosystems but also performs competitively within single-platform contexts.

Applied to posts from Truth Social and X during the 2024 U.S. Presidential election, our framework reveals not only when and where political narratives migrated between ecosystems but also how and through whom this diffusion occurred. Utilizing our framework, we concretely ask:
\begin{enumerate} 
    \item How do political narratives migrate across fragmented platforms like X and Truth Social during the 2024 U.S. Presidential election, and what structural pathways enable their movement?
    
    \item Do shared discourse communities emerge across X and Truth Social during the 2024 election cycle, and if so, how are they positioned within the broader structure of political communication?
    
    \item Are certain users disproportionately associated with cross-platform narrative diffusion during the 2024 U.S. Presidential election, and what structural roles do they occupy within the discourse network?    
\end{enumerate}

Our analysis reveals that political narratives migrated between X and Truth Social in a highly structured manner. Many of these narratives migrated between X and Truth Social through a small but structurally distinct set of users, comprising just 0.33\% of the total. These users accounted for only 2.14\% of all posts, yet served as the initial cross-platform carriers for nearly 70\% of narratives that successfully migrated. These \textit{bridge users} were embedded at the intersection of fragmented discourse communities and consistently appeared early in the life cycle of narratives that later gained traction across ecosystems.

Critically, these insights do not rely on retrospective interaction patterns or explicit behavioral ties. Instead, they emerge from a structural perspective made possible by our modeling approach. By linking users through shared participation in latent discourse clusters, our platform-agnostic framework uncovers connective structures that remain hidden in traditional interaction-based or semantic similarity-based graphs. This shift opens new avenues for studying how influence operates across media environments and positions cross-platform discourse networks as a scalable and generalizable lens for understanding narrative diffusion.




    
        
    

\section{Related Work}

\vspace{3pt} 
\noindent 
\textbf{Network-Based User Representations.} A central challenge in social media analysis is how to represent users in a way that reflects their relationships, ideological alignment, and interactions. Traditional methods rely heavily on platform-specific signals, including follower networks~\cite{bollen2011happiness}, reposts and mentions~\cite{cinelli2021echo}, hashtags~\cite{alieva2022investigating, burghardt2024socio}, or URLs~\cite{ tardelli2024multifaceted}. These signals, however, face increasing limitations due to restrictive data policies~\cite{tromble2021have} and fragmentation of user activities across platforms~\cite{minici2024uncovering}. Semantic approaches, which construct graphs from linguistic similarity of user content~\cite{ng2023coordinating, luceri2024unmasking}, attempt to bypass such limitations. However, these methods typically rely on pairwise comparisons and often suffer from computational complexity and sensitivity to minor linguistic variations~\cite{luceri2024unmasking}. Thus, current methods rarely provide platform-agnostic and scalable solutions.

\vspace{3pt} 
\noindent
\textbf{Behavioral Modeling with Social Graphs.} Once constructed, social graphs are typically evaluated by their utility in downstream tasks, including ideological stance prediction and coordination detection. Ideological stance prediction methods often combine content and network signals~\cite{xiao2020timme, jiang2023retweet}, yet struggle when metadata is sparse or inaccessible~\cite{jiang2023retweet}. Coordination detection generally relies on identifying synchronized behaviors that often depend on behavioral traces that may be limited or inconsistent across platforms ~\cite{luceri2024unmasking, ng2022combined, magelinski2022synchronized}. Overall, existing models are frequently constrained by platform-specific assumptions and data availability, limiting their adaptability in fragmented or low-signal environments.

\vspace{3pt}
\noindent
\textbf{Cross-Platform Dynamics and Behavior.}
As online discourse becomes increasingly fragmented across platforms, users and narratives move fluidly between ecosystems, disrupting traditional modeling assumptions~\cite{ribeiro2021evolution, russo2023spillover}. Understanding these dynamics has become increasingly important, yet modeling them remains methodologically difficult. Cross-platform behavior is hard to capture due to the lack of consistent identifiers and semantic variation across ecosystems~\cite{ng2022cross, ng2023coordinating, luceri2024unmasking, cinus2024exposing, magelinski2022synchronized}. For example, common strategies such as user-matching heuristics~\cite{iofciu2011identifying} and link-tracking~\cite{cinus2024exposing} often struggle with generalizability and scale. Despite growing interest in semantic similarity-based alignment methods, prior approaches remain limited in scope or too narrow to generalize across diverse ecosystems~\cite{luceri2024unmasking, ng2022cross}. Meanwhile, influence operations exploit this fragmentation~\cite{minici2024uncovering, ng2022cross}, underscoring the need for user graph construction methods that are content-driven, platform agnostic, and robust to missing metadata.

\vspace{3pt}
\noindent
\textbf{Structural Influence and Network Theory.}
We connect our concept of \textit{bridge users} to classic theories of structural influence. In sociology and organizational science, \textit{boundary spanners} and \textit{brokers} are actors who enable information flow across otherwise disconnected groups~\cite{cross2002people,gould1989structures}. These roles emphasize that influence stems not only from activity or status, but from an actor’s position within a network. This framing aligns with Granovetter’s theory of weak ties~\cite{granovetter1983strength}, which highlights how loosely connected individuals often facilitate the spread of novel information.

\section{Methods: Platform-Agnostic Social Graphs}

\begin{figure*}[t] 
    \centering 
    \begin{overpic}[width=0.75\textwidth]{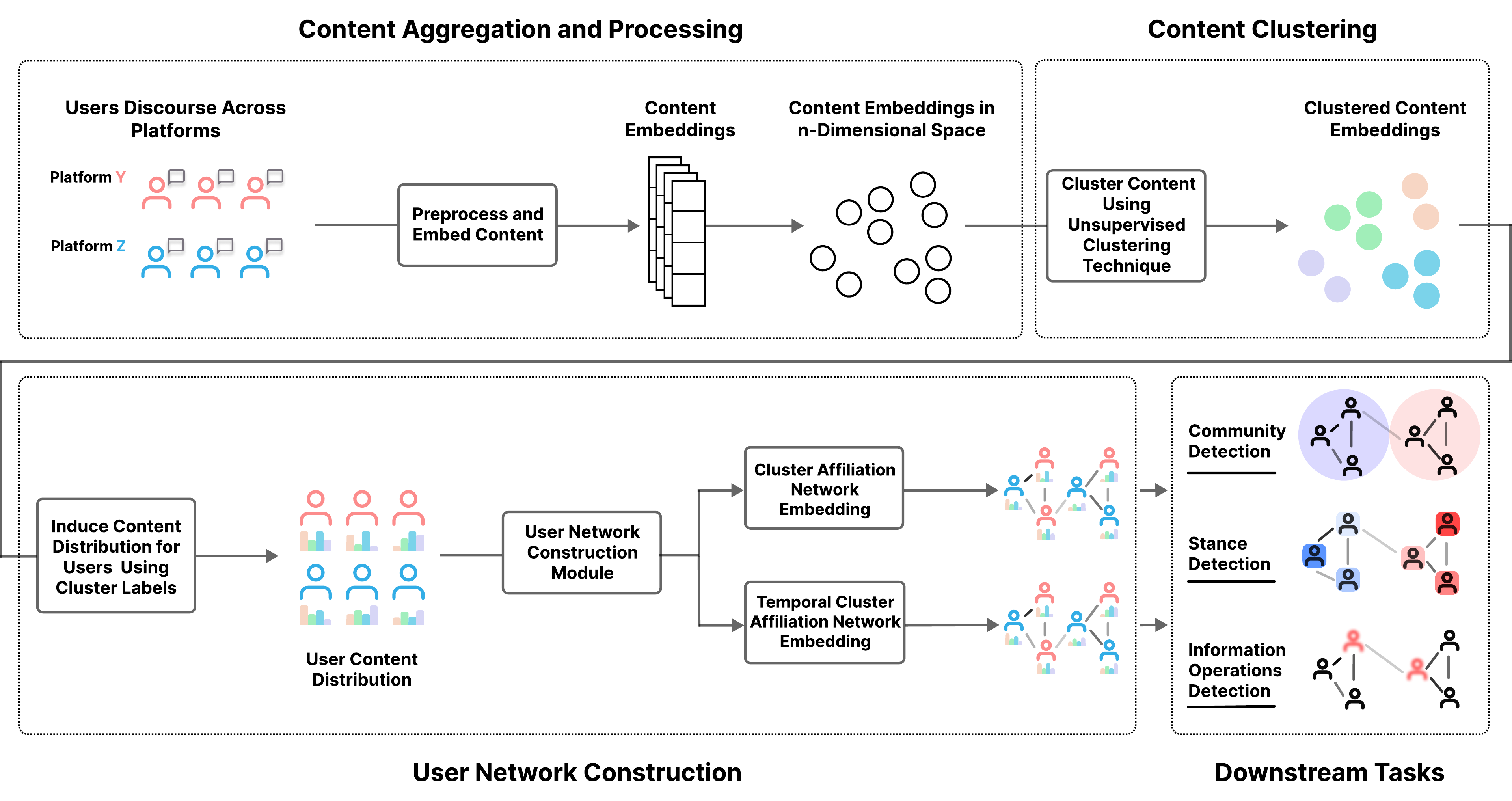}
    \end{overpic}

    \vspace{-1mm} 
    \caption{Overview of our cross-platform user network inference framework. Content is embedded and clustered into semantically coherent narratives, which form the basis for constructing user-user networks. These networks support a range of downstream tasks, including community detection, stance prediction, and content analysis.}
    \vspace{-3mm}
    \label{fig:user-network-generation}
\end{figure*}



Illustrated in Figure~\ref{fig:user-network-generation}, we introduce Cluster Affiliation Network Embedding (CANE): a data-efficient and platform-agnostic framework for constructing user-user networks directly from content. Unlike traditional interaction-based approaches, our method does not depend on reposts, mentions, or other platform-specific signals. Instead, it links users through shared participation in latent narratives, supporting robust graph construction in fragmented, diverse, and sparse data environments. This reframes network modeling around discourse, replacing reliance on platform-specific interactions with the underlying structure of narrative alignment.

Given a set of users $U = {u_1, ..., u_n}$ and their posts $C = {c_1, ..., c_m}$, we construct a weighted, undirected graph $G = (V, E, w)$, where edges capture inferred behavioral alignment via shared participation in semantically coherent content clusters. Our pipeline includes three steps: (1) embedding user posts into a shared semantic space, (2) clustering posts into latent topics, and (3) connecting users based on shared cluster participation. We introduce two variants: \textbf{CANE}, a static model based on aggregated content, and \textbf{t-CANE}, a temporal extension. Full benchmarks and complexity analysis are provided in the Appendix.

\subsection{Content Aggregation and Embedding}

We begin by preprocessing post content for semantic modeling, removing platform-specific artifacts, such as URLs, hashtags, and user mentions. Standard normalization procedures, including whitespace trimming and character encoding cleanup, are applied. We embed each post using MPNet~\cite{song2020mpnet}, a transformer model chosen for its strong performance on semantic similarity tasks. This yields a 768-dimensional vector for each post, which captures contextual meaning while remaining agnostic to platform-specific syntax. We note that our framework is embedding-agnostic and supports substitution with alternative monolingual, multilingual, or multimodal models~\cite{wang2024multilingual}.

\subsection{Content Clustering}
\vspace{3pt}
\noindent
\textbf{DP-Means.}
To identify coherent content clusters, we apply DP-Means to embed representations of user posts. Unlike k-means, which requires specifying the number of clusters in advance, DP-Means adaptively creates new clusters based on a distance threshold $\lambda$. This threshold, which we set using cosine distance and empirical semantic equivalence cutoffs (typically around 0.65~\cite{hanley2023happenstance}), governs when new clusters are formed—allowing us to strike a balance between topical specificity and generality. Full algorithmic details, justification, and sensitivity for different $\lambda$ values are provided in Appendix A.


\subsection{User Network Construction}
Once the content on particular platforms has been clustered (Figure~\ref{fig:user-network-generation}), we utilize two unique strategies for constructing user-user edges, designed to balance expressiveness and computational efficiency. Both methods aim to robustly connect users based on shared narrative engagement, with one capturing static behavioral alignment and the other modeling temporal dynamics.

\vspace{3pt}
\noindent
\textbf{Cluster Affiliation Network Embedding (CANE).} The first method we utilize for constructing user-user edges amongst social media after contracting their participation in different content clusters is based on CANE. CANE models user similarity by capturing how individuals align along content clusters, providing a proxy for their narrative preferences. CANE treats each user's cluster participation as a document-term-like matrix (e.g. akin to methods like TF-IDF~\cite{aizawa2003information}), where clusters are analogous to terms and user participation corresponds to frequency. This formulation allows us to assess not just whether users engage with the same clusters, but how distinctive those shared affiliations are relative to all content clusters.

This is such that for each user $u$, we define its full cluster affiliation across all clusters $C$ vector:
\begin{equation}
   \mathbf{v}_u = [w_{u,1}, w_{u,2}, ..., w_{u,|C|}]
\end{equation}

\noindent
where each weight $w_{u,c}$ for $c \in C$ is computed as:
\begin{equation}
   w_{u,c} = tf_{u,c} \cdot \log\left(\frac{|U|}{|\{u : u \in c\}|}\right)
\end{equation}

Here, $tf_{u,c}$ is the normalized frequency of user $u$'s participation in cluster $c$ (\texttt{i.e}, the number of times that the user posted about the content in the cluster $c$), and the second log term penalizes broadly popular clusters, thereby emphasizing more distinctive affiliations.

Once cluster affiliation vectors are computed, we measure user similarity using cosine similarity, a standard metric for high-dimensional, sparse representations~\cite{song2020mpnet, luceri2024unmasking, pacheco2021uncovering, wang2024multilingual}. This captures the proportional overlap in narrative engagement, ensuring that users with similarly distributed content preferences are more strongly connected, even if they do not share identical cluster memberships. We evaluated alternative weighting strategies, including raw counts and softmax normalization, and found that a TF-IDF-inspired scheme yielded the best performance in downstream tasks. This supports the intuition that TF-IDF weighting better emphasizes distinctive cluster participation while down-weighting generic, viral, and ubiquitous content (Appendix B; Table~\ref{tab:cane_weighting_ablation}).

To improve scalability, we compute user-user similarity using the GPU-supported FAISS\footnote{\url{https://github.com/facebookresearch/faiss}} library~\cite{malkov2018efficient}. This approximate nearest neighbor method replaces brute-force comparisons, supports efficient edge construction, and produces sparse, high-quality user graphs. Empirical evaluation on multiple datasets shows that using FAISS-HNSW yields performance comparable to brute-force similarity computation, with no statistically significant degradation in F1 or AUC (Appendix B; Table~\ref{tab:cane_faiss_eval}).

\vspace{3pt}
\noindent
\textbf{Temporal Cluster Affiliation Network Embedding (t-CANE).} While the static CANE model effectively captures user alignment through shared narrative engagement, it collapses time, obscuring the dynamics of when and how relationships form and shift. Yet many critical processes in political discourse, such as coordination, radicalization, and narrative seeding, are inherently temporal \cite{tardelli2024temporal}. To capture these dynamics, we introduce \textbf{t-CANE}, a discrete temporal extension. This allows us to trace how user affinities evolve and to distinguish fleeting overlaps from sustained ideological alignment.

At each timestep \( t \), we compute a TF-IDF-weighted cluster affiliation vector \( \mathbf{v}_u^t \) for each user \( u \), summarizing their narrative participation during that window. Pairwise similarities are measured via cosine similarity. To construct a dynamic graph, we apply a memory-based update rule inspired by Hawkes processes~\cite{laub2015hawkes}, which reinforces repeated user alignment across timesteps while gradually weakening connections that lapse. This mechanism allows the graph to capture both persistent relationships and newly emerging discourse patterns. Neighborhoods at each timestep are computed using the GPU-supported FAISS library to ensure scalable similarity search. We validate this design through downstream task performance and targeted ablation studies (Appendix~C).






\section{Benchmark Tasks and Evaluation}
We evaluate our framework on three tasks: information operation detection, ideological stance prediction, and a new cross-platform engagement prediction benchmark. The first two tasks test intra-platform utility, while the third assesses whether user-user networks capture behavioral alignment across ecosystems. For each task, we compare the user-user network construction methods in Table~\ref{tab:similarity_networks_summary} against our own while keeping the downstream prediction model fixed. This setup thus isolates the input graph's role on performance. 


To assess robustness under sparse data conditions, we simulate increasing data availability by incrementally sampling $n\%$ of each user’s posts in 5\% steps, retaining previously included posts for consistency. As part of our evaluation framework, we define the point of data efficiency of a method as the point at which it reaches 95\% of its maximum AUC. This threshold serves as a standard for all methods and offers a principled balance between performance and efficiency. We validate the appropriateness of this choice by confirming that performance increases smoothly with additional data, and that the 95\% point is not an outlier but reflects consistent trends (see Figure~\ref{fig:peak-auc}).

\begin{table}[t]
    \centering
    \small
    \renewcommand{\arraystretch}{1.25}
    \resizebox{0.49\textwidth}{!}{ 
    \begin{tabular}{@{}lp{7cm}@{}}
        \toprule
        \textbf{Network Type} & \textbf{Construction Description} \\
        \midrule
        Co-Repost & Connect users who repost the same piece of content. \\
        \midrule
        Co-URL & Connect users who share the same URLs in their posts. \\
        \midrule
        Fast Repost & Connect users who repost identical content within a short time window. \\
        \midrule
        Hashtag Sequence & Connect users based on ordered sequences of shared hashtags. \\
        \midrule
        Text Similarity & Connect users if they post at least one highly similar post; edge weight reflects average text similarity across matches~\cite{pacheco2021uncovering}. \\
        \midrule
        k-NN Embedding Graph & Build a full text-to-text kNN graph first, then induce a user-to-user graph reflecting overall proximity across all posts~\cite{ng2023coordinating}. \\
        \midrule
        Fused Graph & Construct a unified network where users are linked if they are connected in any underlying similarity network (Co-Repost, Co-URL, Fast Repost, Hashtag Sequence, or Text Similarity)~\cite{luceri2024unmasking}. \\
        \bottomrule
    \end{tabular}
    }
    \caption{Baseline similarity network construction methods.}
    \label{tab:similarity_networks_summary}
\end{table}

\begin{table}[t]
\small
\centering
\renewcommand{\arraystretch}{1.1}
\resizebox{0.4\textwidth}{!}{
\begin{tabular}{lcc}
\toprule
\textbf{Method} & $\text{Macro-F}_1$ & AUC \\
\midrule
Co-Repost & 0.71 $\pm$ 0.02 & 0.77 $\pm$ 0.03 \\
Co-URL & 0.50 $\pm$ 0.02 & 0.56 $\pm$ 0.06 \\
Fast Repost & 0.46 $\pm$ 0.01 & 0.54 $\pm$ 0.02 \\
Hashtag Sequence & 0.51 $\pm$ 0.04 & 0.49 $\pm$ 0.03 \\
Text Similarity & 0.53 $\pm$ 0.01 & 0.33 $\pm$ 0.00 \\
k-NN Embedding Graph & 0.58 $\pm$ 0.02 & 0.39 $\pm$ 0.01 \\
Fused Graph & 0.74 $\pm$ 0.02 & 0.81 $\pm$ 0.02 \\
CANE & 0.72 $\pm$ 0.02 & 0.91 $\pm$ 0.01 \\
t-CANE & \textbf{0.83 $\pm$ 0.01} & \textbf{0.98 $\pm$ 0.01} \\
\bottomrule
\end{tabular}
}
\caption{Performance on the China IO dataset.}
\label{tab:china_io}
\vspace{-3mm}
\end{table}

\begin{table}[t]
\small
\centering
\renewcommand{\arraystretch}{1.1}
\resizebox{0.4\textwidth}{!}{
\begin{tabular}{lcc}
\toprule
\textbf{Method} & $\text{Macro-F}_1$ & AUC \\
\midrule
Co-Repost & 0.80 $\pm$ 0.01 & 0.73 $\pm$ 0.01 \\
Co-URL & 0.49 $\pm$ 0.00 & 0.52 $\pm$ 0.01 \\
Fast Repost & 0.59 $\pm$ 0.01 & 0.60 $\pm$ 0.01 \\
Hashtag Sequence & 0.54 $\pm$ 0.03 & 0.48 $\pm$ 0.01 \\
Text Similarity & 0.62 $\pm$ 0.03 & 0.52 $\pm$ 0.02 \\
k-NN Embedding Graph & 0.65 $\pm$ 0.01 & 0.57 $\pm$ 0.00 \\
Fused Graph & 0.81 $\pm$ 0.01 & 0.76 $\pm$ 0.01 \\
CANE & 0.82 $\pm$ 0.01 & 0.94 $\pm$ 0.00 \\
t-CANE & \textbf{0.90 $\pm$ 0.01} & \textbf{0.94 $\pm$ 0.02} \\
\bottomrule
\end{tabular}
}
\caption{Performance on the Iran IO dataset.}
\label{tab:iran_io}
\vspace{-3mm}
\end{table}

\subsection{Information Operations}
Information operations (IOs) represent a critical and high-stakes challenge for online platform governance and political communication research. Detecting such operations often requires robust inference of latent user alignment in the absence of explicit coordination signals, making them a strong test case for evaluating the effectiveness of user-user network construction methods. We evaluate our framework on the task of IO detection using the dataset from~\citet{seckin2024labeled}, which includes labeled state-backed campaigns identified by Twitter and released to the research community to enable research on IO detection. We focus on two operations: one attributed to Iran and one to China, each comprising users labeled as either IO drivers or organic accounts. See Appendix D for details.

Following the methodology of~\citet{luceri2024unmasking}, we embed user similarity networks (Table~\ref{tab:similarity_networks_summary}) where edges reflect inferred alignment between users based on shared content or behavioral patterns using node2vec~\cite{grover2016node2vec} and train a random forest classifier to distinguish drivers from organic users.

\vspace{2pt}
\noindent
\textbf{Overall Results.}
Shown in Tables~\ref{tab:china_io} and~\ref{tab:iran_io}, our method achieves the highest performance across both campaigns, with t-CANE reaching 0.85 $\text{Macro-F}_1$ and 0.92 AUC on the China IO dataset, and 0.90 $\text{Macro-F}_1$ and 0.94 AUC on the Iran IO dataset. The temporal variant (t-CANE) consistently outperforms static methods, highlighting the importance of modeling evolving behavior.

\vspace{2pt}
\noindent
\textbf{Sparse Data Setting.}
To evaluate data efficiency, we incrementally sample user content in 5\% intervals and report how much data each method requires to reach 95\% of its peak AUC. Our method achieves near-optimal performance with just 5–10\% of the training data, significantly outperforming baselines. Although semantic similarity-based methods (e.g., k-NN Embedding Graph) also show data efficiency, they plateau at much lower overall performance. Full results are available in Appendix D (Table~\ref{tab:data_efficiency_95auc_IO}).

\subsection{Ideological Mapping on X and TikTok}
\label{sec:ideology-classification}
We apply our framework to an ideological classification task using 2024 U.S. election data from X and TikTok, sampled via a shared keyword query~\cite{pinto2024tracking2024presidentialelection, balasubramanian2024public}. The task involves predicting each user’s political leaning as either Liberal or Conservative, a common binary framing in prior work on ideology prediction~\cite{jiang2023retweet}. X serves as a standard benchmark for this task; TikTok presents a more challenging and increasingly common environment, where traditional graph construction is constrained by sparse metadata and platform limitations, making it a useful testbed for evaluating structure-based approaches.

Many ideological stance prediction models rely on the structure of user-user graphs, often constructed from follower links or interaction data that reflect homophily and alignment~\cite{xiao2020timme, jiang2023retweet}. To isolate the role of structure, we exclude all content features and apply a fixed downstream model across all graph variants. This setup allows us to evaluate how well each network captures latent ideological patterns, independent of content or platform-specific behaviors.

As in previous sections, we compare against the baseline network construction methods listed in Table~\ref{tab:similarity_networks_summary}. On TikTok, where available signals are limited, we evaluate Hashtag Sequence, Text Similarity (from video descriptions), k-NN embedding Graph, and a partial fused graph.

We evaluate classification performance using graph-based models. While multiple architectures were tested, final results are reported using node2vec embeddings combined with a feedforward neural network classifier. This setup emphasizes structural information, aligning with our goal of isolating graph quality without relying on content features; performance was comparable across alternatives. See Appendix~F for details. 

\vspace{2mm}
\noindent
\textbf{US-2024 X Dataset.} We use the large-scale dataset of political discourse on X during the 2024 U.S. election introduced by~\citet{balasubramanian2024public}. From this, we sample 24,000 users who posted at least five times. This threshold filters out the bottom 5\% of users, who are often associated with low-quality or noise-prone content producers, and aligns with standard practices for reducing volatility in downstream modeling~\cite{luceri2024susceptibility}. A manually labeled subset of 2,100 users was annotated as liberal, conservative, or other/NA. Labeling was conducted by trained annotators with political science and media studies expertise, following standardized coding guidelines~\ref{tab:coding-guidelines}. Each user was independently labeled by two raters; disagreements were excluded from the final dataset. See Appendix E for full label distributions and inter-annotator agreement statistics.

\vspace{2mm}
\noindent
\textbf{US-2024 TikTok Dataset.} We use the corresponding 2024 U.S. election dataset on TikTok from~\citet{pinto2024tracking2024presidentialelection}, sampling 52,859 active users ($\geq$ 5 posts). A subset of 1,758 users was manually labeled using both video descriptions and Whisper-transcribed audio
\footnote{\url{https://github.com/openai/whisper}}, following the same schema as for the X dataset. Labeling metrics for both platforms appear in Appendix E (Table~\ref{tab:annotation-stats-twitter-tiktok}).

\vspace{3pt}
\noindent
\textbf{US-2024 X Results.} As shown in Table~\ref{tab:similarity_networks-twitter-political}, our method achieves the highest $\text{Macro-F}_1$ (0.83) and AUC (0.76) scores across all network construction approaches, demonstrating strong downstream performance. Traditional networks underperform largely due to platform-specific signal sparsity. For instance, only 0.003\% of X posts contain URLs, and just 0.65\% of users share them—rendering URL-based graphs nearly empty. Repost and hashtag graphs perform similarly, covering only 5.2\% and 5.4\% of users, respectively. Even the Fused Graph struggles with limited behavioral coverage.

To isolate performance from coverage effects, we evaluate each method using only the users included in its graph. As shown in Table~\ref{tab:method-vs-network}, our method remains competitive under these constraints, suggesting its advantage stems from structural quality, not just breadth. Class distributions and coverage-adjusted subsets appear in Table~\ref{tab:annotation-stats-controlled}.

\vspace{3pt}
\noindent
\textbf{US-2024 TikTok Results.} As shown in Table~\ref{tab:similarity_networks-tiktok-political}, our method again achieves the highest performance ($\text{Macro-F}_1$: 0.83, AUC: 0.83). Unlike the X-based evaluation, no user filtering is needed here, as the graph covers 100\% of labeled users, likely due to the dataset design by~\citet{pinto2024tracking2024presidentialelection}, where hashtags appear in 92.6\% of video descriptions.

\vspace{3pt}
\noindent
\textbf{Sparse Data Setting: US-2024 X and TikTok Results.} We observe similar trends in the US-2024 X and TikTok datasets as in the IO experiments: our networks require significantly less data to reach near-optimal performance. As shown in Table~\ref{tab:data_efficiency_95auc_IO-ideological}, our methods reach 95\% of their peak AUC with minimal user content (5-10\%), consistently outperforming alternatives in both efficiency and final performance. Likely due to their operation on individual posts, semantic similarity-based methods also exhibit relatively low data requirements. However, consistent with our IO  setting findings, these methods plateau at lower performance levels compared to our method.

\begin{table}[th]
    \centering
    \renewcommand{\arraystretch}{1}
    \resizebox{0.42\textwidth}{!}{
    \begin{tabular}{llcc}
        \toprule
        \textbf{Method} & \textbf{Similarity Network} & \textbf{$\text{Macro-F}_1$} & \textbf{AUC} \\
        \midrule
        Baseline & Co-Repost & 0.43 $\pm$ 0.30 & 0.56 $\pm$ 0.22 \\
        & Co-URL & 0.51 $\pm$ 0.26 & 0.53 $\pm$ 0.05 \\
        & Fast Repost & 0.48 $\pm$ 0.04 & 0.51 $\pm$ 0.03 \\
        & Hashtag Sequence & 0.65 $\pm$ 0.22 & 0.63 $\pm$ 0.14 \\
        & Text Similarity & 0.39 $\pm$ 0.12 & 0.32 $\pm$ 0.09 \\
        & k-NN Embedding Graph & 0.41 $\pm$ 0.03 & 0.34 $\pm$ 0.11\\
        & Fused Graph & 0.69 $\pm$ 0.15 & 0.70 $\pm$ 0.05\\
        \midrule
        \multirow{3}{*}{Ours} 
        & CANE & 0.81 $\pm$ 0.01 & 0.73 $\pm$ 0.02 \\
        & t-CANE &\textbf{ 0.83 $\pm$ 0.01} & \textbf{0.76 $\pm$ 0.02} \\
        \bottomrule
    \end{tabular}}
        \caption{Ideological classification on X.}
    \label{tab:similarity_networks-twitter-political}
\end{table}
\vspace{-3mm}

\begin{table}[th]
    \centering
    \renewcommand{\arraystretch}{1}
    \resizebox{0.42\textwidth}{!}{
    \begin{tabular}{llcc}
        \toprule
        \textbf{Method} & \textbf{Similarity Network} & \textbf{$\text{Macro-F}_1$} & \textbf{AUC} \\
        \midrule
        Baseline
        & Hashtag Sequence & 0.71 $\pm$ 0.03 & 0.62 $\pm$ 0.01 \\
        & Text Similarity & 0.38 $\pm$ 0.14 & 0.42 $\pm$ 0.07 \\
        & k-NN Embedding Graph & 0.45 $\pm$ 0.10 & 0.45 $\pm$ 0.08 \\
        & (Partial) Fused Graph & 0.71 $\pm$ 0.03 & 0.64 $\pm$ 0.02 \\
        \midrule
        Ours & CANE & \textbf{0.83 $\pm$ 0.02} & \textbf{0.83 $\pm$ 0.01} \\
        & t-CANE & 0.82 $\pm$ 0.02 & \textbf{0.83 $\pm$ 0.03} \\
        \bottomrule
    \end{tabular}}
        \caption{Ideological classification on TikTok.}
    \label{tab:similarity_networks-tiktok-political}
        \vspace{-3mm}
\end{table}

\subsection{Cross-Platform Engagement Prediction}
We introduce a novel benchmark to test whether user-user networks can predict future topic engagement across platforms. This evaluates whether structural proximity in the graph corresponds to behavioral alignment in fragmented media ecosystems. Formulated as a user-topic prediction task, the model receives a graph at time $t$ and predicts if user $u$ will engage with topic $k$ in the future using only structural features. No content-based features are used during prediction, isolating the representational value of the graph.

We apply this to a dataset of U.S. election-related posts from X and Truth Social (May–Nov 2024), clustered into 321 cross-platform narrative themes using multilingual MPNet embeddings and DP-means clustering. Each theme is labeled using top TF-IDF terms. Full preprocessing details, cluster labeling procedures, and representative narrative examples are provided in Appendix~G. As in prior tasks, we train GCNs~\cite{zhang2019graph} on each graph, using binary user-topic features to predict future engagement. We report performance at multiple time windows using Macro-$\text{F}_1$ and AUC. A randomly rewired graph is included to verify that performance reflects real structure, not model capacity.

\begin{table}[th]
    \centering
    \renewcommand{\arraystretch}{1}
    \resizebox{0.42\textwidth}{!}{
    \begin{tabular}{llll}
        \toprule
        \textbf{Method} & \textbf{Similarity Network} & \textbf{$\text{Macro-F}_1$} & \textbf{AUC} \\
        \midrule
        Baseline
        & Random GCN & 0.00 $\pm$ 0.00 & 0.56 $\pm$ 0.03 \\
        & Co-URL & 0.01 $\pm$ 0.00 & 0.43 $\pm$ 0.04 \\
        & Hashtag Sequence & 0.11 $\pm$ 0.04 & 0.48 $\pm$ 0.08 \\
        & Text Similarity & 0.02 $\pm$ 0.00 & 0.58 $\pm$ 0.05 \\
        & k-NN Embedding Graph & 0.02 $\pm$ 0.01 & 0.61 $\pm$ 0.06 \\
        & (Partial) Fused Graph & 0.05 $\pm$ 0.01 & 0.64 $\pm$ 0.06 \\
        \midrule
        Ours & CANE & 0.30 $\pm$ 0.02 & 0.89 $\pm$ 0.02 \\
        & t-CANE & \textbf{0.35 $\pm$ 0.06} & \textbf{0.94 $\pm$ 0.02} \\
        \bottomrule
    \end{tabular}}
        \caption{Performance of similarity networks for cross-platform narrative engagement prediction (t=7).}
    \label{tab:similarity_networks-cross-platform}
        \vspace{-3mm}
\end{table}

\vspace{3pt}
\noindent
\textbf{Results.}
 As shown in Table~\ref{tab:similarity_networks-cross-platform} and corroborated across all timesteps in Appendix Table~\ref{tab:full-engagement-prediction}, our proposed approach, t-CANE, substantially outperforms all baseline methods. The strongest baseline achieves a $\text{Macro-F}_1$ score of 0.11 (Hashtag Sequence), while t-CANE reaches 0.35, more than tripling predictive accuracy.
 Even using our non-temporal model CANE, our model reached a $\text{Macro-F}_1$ of 0.30. Similarly, the highest baseline AUC is 0.64 (Fused Graph), compared to 0.94 for t-CANE.


\section{Connective Narratives: X and Truth Social}

To analyze how political narratives migrated between Truth Social and X during the 2024 U.S. Presidential election, we use the CANE framework to construct a unified discourse network that connects users through shared narrative engagement. This representation allows us to investigate not only whether narratives moved between platforms, but also how, when, and through whom they traveled. Our analysis proceeds in three stages: first, we characterize overall patterns of narrative migration; second, we model the structure of narrative flow to assess whether it follows concentrated and repeatable routes; and third, we examine whether a small set of users embedded across discourse communities consistently enables cross-platform diffusion.

We find that narrative diffusion between X and Truth Social followed structured and repeatable patterns. It was shaped by asymmetries in platform roles and by repeated involvement from users positioned at the boundaries of discourse communities. By viewing the ecosystem through a discourse network, we are able to uncover concrete pathways through which narratives consistently migrated, pathways that remain invisible to interaction- or semantic similarity-based. In the sections that follow, we use this structural lens to characterize patterns of narrative movement, identify the users who drive it, and assess the broader architecture of the cross-platform discourse that occurred.


\subsection{Cross-Platform Narrative Migration}

To understand how political discourse flows across platforms, we begin by modeling narratives, seeking to trace how coherent narratives emerge, evolve, and migrate between Truth Social and X. Following prior work on narrative tracking in multilingual information ecosystems~\cite{hanley2024specious}, we define a narrative as a collection of posts that focus on the same issue or event. This definition treats narratives as coherent units of discourse that evolve over time and platform, rather than as isolated posts or terms. For example, in the context of the 2024 U.S. election, a narrative might involve claims that mail-in ballots are being used to rig the outcome: potentially unfolding over multiple posts that cite anecdotal evidence, inaccurate information, and emotionally charged appeals.

To identify narratives, we embed posts using the multilingual MPNet model fine-tuned for semantic similarity and apply DP-means clustering with a cosine distance cutoff of 0.30 (i.e., minimum similarity of 0.70 within clusters). Prior work in narrative tracking~\cite{hanley2024specious} typically uses thresholds between 0.60 and 0.80. Given the greater linguistic variability across Truth Social and X, we adopt a 0.65 threshold to balance semantic coherence with topical specificity, minimizing the risk of merging distinct narratives. We validate this choice via human evaluation of 50 post pairs at various thresholds (Table~\ref{tab:threshold-accuracy-method}) and find this threshold to be comparable to benchmarks in similar work~\cite{hanley2024partial}.


To establish a baseline for narrative movement, we first identify cases of \textit{simple migration}: narratives that cross platform boundaries, regardless of sustained dynamics. A narrative is considered to originate on a platform if it appears there at least 24 hours before surfacing on the other, filtering out near-simultaneous mentions triggered by shared external events. Migration is confirmed when the narrative surpasses a minimum engagement threshold on the receiving platform: defined as 10 posts, the 35th percentile across all narrative-platform pairs. This ensures focus on substantive diffusion events and follows best practices in prior coordination studies~\cite{magelinski2022synchronized, nizzoli2021coordinated}. We validate this threshold via sensitivity analysis (Table~\ref{tab:threshold-accuracy-method}).

Next, to assess structured and sustained diffusion, we identify cases of \textit{significant migration}: narratives showing statistically directional flow. We compute Transfer Entropy (TE) to assess directional information flow. TE is a non-linear, asymmetric measure that detects whether past activity on one platform reliably reduces uncertainty about future activity on the other~\cite{schreiber2000measuring}, allowing us to move beyond simple symmetric association to capture predictive, time-structured diffusion dynamics. To validate directionality, we perform a permutation test by randomly shuffling the source platform’s time series 1,000 times to generate a null TE distribution. Narratives are considered significant if their observed TE exceeds the 95th percentile of this distribution ($p < 0.05$). This procedure isolates narratives with robust, predictive diffusion dynamics.

\vspace{3pt}
\noindent
\textbf{Results.} From this process, we identify 1,552 narratives exhibiting at least \textit{simple migration} and 238 narratives exhibiting directional, \textit{significant migration} (that is, 15.3\% of narratives exhibiting at least simple migration also exhibit significant migration). We enumerate example representative narratives along with their TE in Table~\ref{tab:te-narratives}.

At the level of \textit{simple migration}, we find that Truth Social plays a disproportionately large role in initiating cross-platform diffusion. Despite accounting for just 1.7\% of total post volume across migrating narratives, Truth Social is the origin point for 18.9\% of them, making its narratives more than eleven times more likely to initiate cross platform spread than expected by chance. These findings suggest that Truth Social, although small in scale, may serve as an important incubator for narratives that later gain broader traction.

Qualitatively, narratives originating on Truth Social tend to exhibit more overt themes of fear or conspiratorial framing, often revolving around election integrity, government control, or existential threat. This contrasts with X originating narratives, which more commonly emphasize partisan conflict or elite criticism without the same fear-centric framing. To test this distinction systematically, we apply a model fine-tuned on human-annotated fear speech data~\cite{saha2023rise} to all posts within each migrating narrative (see Appendix~H for details). Narratives seeded from Truth Social exhibit significantly higher rates of fear-laden language than those originating on X ($p < 0.01$), with a log-odds ratio of +0.22 and a 22.5\% relative increase.


Turning to \textit{significant migration}, our stricter criteria yield 238 narratives exhibiting statistically significant, directional cross-platform flow. These narratives form the empirical foundation for the rest of our analysis. Within this set, we again find that Truth Social plays an outsized role in shaping subsequent activity on X. Despite contributing only 1.1\% of total post volume within these clusters, Truth Social leads 15.6\% of significant migration cases, with the remainder originating from X. representing a 14.2$\times$ overrepresentation relative to its share of total post volume (1.1\%), calculated as the ratio between observed and expected initiation rates. This suggests that Truth Social frequently functions as an upstream source within the temporal structure of cross-platform discourse. Overall, these results indicate that the observed cross-platform migration is common, structurally directional, and disproportionately seeded by Truth Social.

\vspace{-3pt}
\subsection{Discourse as Structures: Bridge Zones}

To examine the structural dynamics of the narrative flow between X and Truth Social, we adopt a discourse-centered approach. Rather than treating each platform as an isolated system, we use CANE to model a unified information ecosystem, where users are linked through
shared narrative participation. To uncover its internal organization, we apply the Louvain algorithm~\cite{traag2019louvain} to detect discourse communities: groups of users who consistently engage with similar narratives over time. We then assess the cross-platform composition of each community using Shannon entropy, a standard measure of distributional diversity. Here, entropy reflects how evenly users are distributed across platforms: values near 0 indicate platform homogeneity, while values near 1 suggest a balanced mix.

\vspace{3pt}
\noindent
\textbf{Bridge Communities.} 
We identify a single high-entropy community (entropy $= 0.72$) comprising just 0.33\% of users and 2.14\% of posts (visualized in Figure~\ref{fig:bridge-zone-visualization}). Notably, this community's users do not stand out in terms of volume, but through their distinctive position within the discourse network inferred from content alignment. They span discursive boundaries between platforms, occupying a region through which narratives are especially likely to migrate.

\begin{figure}[t]
    \centering
    \includegraphics[width=0.45\textwidth]{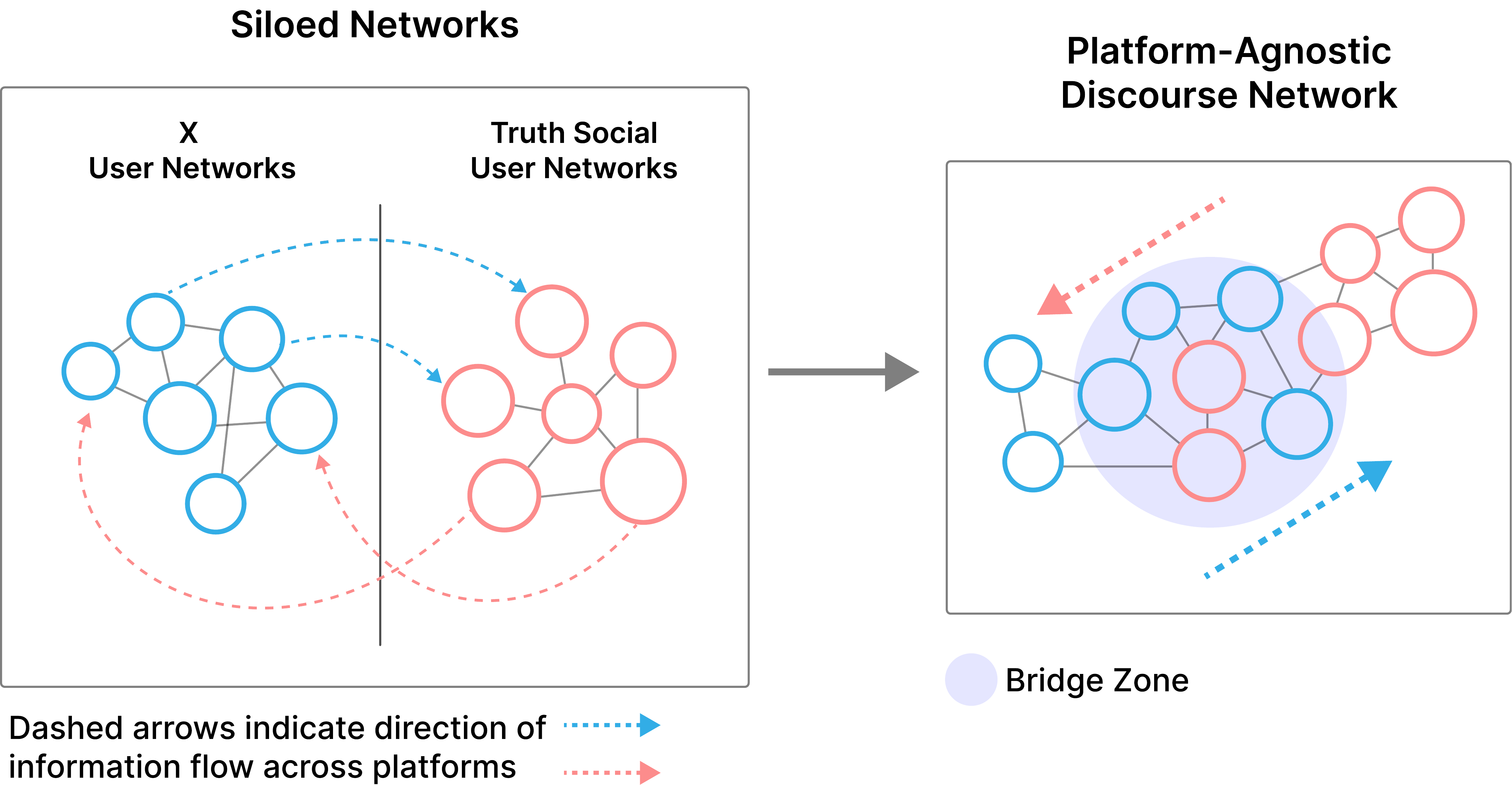}
    \caption{
    A conceptual illustration of cross-platform narrative diffusion. While most communities are siloed within platforms, some users form \textit{bridge zones}: structural overlaps where narrative transfer is more likely.
    }
    \vspace{-3mm}
    \label{fig:bridge-zone}
\vspace{-2mm}
\end{figure}

We refer to this region as a \textit{bridge zone}: a conceptual area in the discourse network where users from different platforms are densely connected through shared narrative engagement (illustrated in Figure~\ref{fig:bridge-zone})~\cite{mendelsohn2023bridging}. To validate that this zone reflects meaningful structural alignment rather than artifacts of duplicated accounts across platforms, we apply a two-stage de-duplication process across the full user set. First, we compute character-level string similarity using the Ratcliff/Obershelp algorithm~\cite{hasugian2023image}, removing 1,644 high-similarity username pairs (0.19\% of users) based on a liberal threshold of 0.7. Second, to assess potential residual duplication, we manually audit 200 randomly sampled users from the \textit{bridge zone}. For each, we searched the other platform’s user interface for accounts with matching or near-matching usernames and names where applicable. In cases of near matches, we cross-referenced bios and profile images to assess the likelihood of duplication. Only six users (3\%) appeared to maintain accounts on both platforms, and none of them were retained in the final data set. This suggests that account duplication is unlikely to explain the distinct structural role of the observed \textit{bridge zone}.


Importantly, these structural dynamics only become visible when discourse is modeled as a unified, cross-platform network. While some baseline graph construction methods surface large or mildly cross-platform communities, none isolate a region with such substantial platform overlap and strong association with narrative migration. See Appendix~I for empirical comparisons.


\definecolor{our_blue}{HTML}{0075F2}

\vspace{3pt}
\noindent

\noindent
\textbf{Bridge Users.} Building on the concept of bridge zones (regions of structural overlap in the discourse network where narratives migrate across platforms), we define \textit{bridge users} as members of these high-entropy, cross-platform communities. Rather than standing out by volume or engagement, they are structurally embedded at key intersections between Truth Social and X. Notably, their activity metrics are near the global median (Appendix I, Table~\ref{tab:bridge_user_activity}), suggesting their influence stems from position, not popularity.

These 2,864 \textit{bridge users} comprise just 0.33\% of all users and are responsible for only 2.14\% of posts, yet they play a disproportionately large role in cross-platform diffusion. They serve as the first cross-platform carriers for 68\% of simple migrating narratives and 69\% of significant migrating narratives: an overrepresentation of more than 200$\times$ their population share. In addition, they are not only conduits, but also initiators, seeding 26\% of all narratives, including 22\% of simple migrating narratives and 19\% of significant migrating narratives.


%

Interestingly, even within this small subgroup, influence is highly concentrated. Just 122 bridge users account for all earliest introductions of Truth Social-originating narratives into X, with four users alone responsible for roughly a quarter of these events. Although we claim no evidence of explicit coordination, two of these users independently pinned an identical inflammatory image (“F*** this Government”) nearly a year apart, pointing to convergent symbolic signaling despite the absence of direct ties. These findings suggest that bridge users occupy structurally important points in the discourse network: locations where cross-platform narrative movement is significantly more likely to occur. 




\vspace{3pt}
\noindent
\textbf{Bridge Users and Narrative Diffusion.} As shown above, \textit{bridge users} played a disproportionately large role in the narrative flow. However, visibility alone does not explain their significance. Structurally embedded actors who engage early can play a decisive role in shaping how information spreads, particularly in fragmented systems where direct ties are sparse~\cite{granovetter1983strength, gould1989structures}. Engagement at the early stages of a narrative's lifecycle reflects not just activity but also a form of narrative buy-in which may determine which ideas take root and diffuse. We therefore test whether narratives in which bridge users appear during the earliest stages of formation (defined as within the first 5\% of participants; we validate the robustness of this in Appendix~I) exhibit higher downstream engagement than those initiated by comparable users.

To ensure a fair comparison, each bridge user is matched to a non-bridge user from a single-platform community using k-nearest neighbors based on z-normalized behavioral metrics: post volume, likes, replies, and reposts. We then evaluate engagement across likes, replies, and reposts. To quantify the strength of the observed differences, we report the rank-biserial correlation, which measures the probability that a randomly selected value from one group will exceed one from the other, providing an interpretable effect size for non-parametric comparisons. Narratives involving early bridge user participation receive significantly more interaction across all metrics (Mann–Whitney $U$, $p < 0.01$), with large effect sizes for likes (0.60) and reposts (0.46), and a moderate effect for replies (0.34). These results are robust across a range of early engagement thresholds (see Appendix~I), and hold even when restricting to low-virality narratives, indicating that bridge users are not simply reacting to already popular content.

These patterns may bolster the idea that these \textit{bridge users} partially function as early validators structurally positioned at the seams of the discourse network. Their engagement aligns with classic network theory, where actors such as brokers~\cite{gould1989structures} and weak ties~\cite{granovetter1983strength} help facilitate early information flow across community boundaries. Rather than reflecting visibility or volume, their engagement patterns appear to coincide with structural embedding in positions that are more likely to be associated with cross-platform diffusion.

\vspace{-3pt}

\begin{table}[t]
\centering
\resizebox{1.03\linewidth}{!}{%
\begin{tabular}{lc}
\toprule
\textbf{Narrative Theme} & \textbf{TE (TS $\rightarrow$ X)} \\
\midrule
Criticism of Stormy Daniels' testimony against Trump. & 0.95 \\
Discussions around Bud Light's sales slump after boycott fallout. & 0.58 \\
Claims about FBI suppression regarding Seth Rich. & 0.50 \\
Allegations of FBI suppression of the Hunter Biden laptop story. & 0.49 \\
Posts suggesting Biden mental decline. & 0.49 \\
Valorization of El Salvador President, Nayib Bukele. & 0.36 \\
Accusations of Kamala Harris being racial ``chameleon.'' & 0.33 \\
Calls to ``cancel'' people, events, or shows. & 0.31 \\

\bottomrule
\end{tabular}
}
\caption{Representative narratives exhibiting significant directional influence from Truth Social to X, as identified via Transfer Entropy. Themes were assigned by a human expert based on posts and top TF-IDF terms within each narrative.}
\label{tab:te-narratives}
\vspace{-5mm}
\end{table}

\section{Discussion and Conclusion}

Political narratives during the 2024 U.S. Presidential election did not diffuse haphazardly across X and Truth Social. Rather, they followed structured and asymmetric paths, enabled by a small subset of users embedded at the intersection of fragmented discourse communities. These \textit{bridge users}, who comprise just 0.33\% of users and account for only 2.14\% of posts, nonetheless served as initial cross-platform conduits for approximately 70\% of narratives that migrated between platforms. Their centrality to cross-platform narrative flow stemmed not from visibility or volume, but from their structural embedding within a unified discourse network that spanned both ecosystems. Crucially, this structure only becomes visible when discourse is modeled directly. While some baseline graph methods surface modest cross-platform communities, none recover the tightly embedded bridge zone that emerges from our discourse network. This not only allows us to recover patterns of cross-platform diffusion that remain hidden in interaction-based graphs, but also makes our method scalable and robust in environments where behavioral data is limited or siloed.

These results expand our understanding of influence in fragmented media. Rather than treating platforms as silos, our approach surfaces the connective tissue through which narratives moved across X and Truth Social. This highlights the unique value of discourse-centered modeling: it not only improves prediction but also reveals latent structures of narrative diffusion that remain hidden in traditional, behavior- or metadata-based graphs. Moreover, this structural view aligns with classic theories in sociology and organizational science. Our notion of ``bridge users'' parallels boundary spanners in organizational networks~\cite{cross2002people} and brokers in Gould and Fernandez’s typology~\cite{gould1989structures}. Most directly, it aligns with Granovetter’s theory of weak ties~\cite{granovetter1983strength}, which highlights how those positioned across structural holes, rather than central figures, often drive information flow. In our case, bridge users span platform divides, occupying boundary positions in the discourse network that are consistently associated with cross-platform diffusion.

As such, our contributions are both empirical and methodological. Empirically, we show that narrative diffusion across X and Truth Social exhibits highly structured patterns, with a small set of users consistently positioned at key points of cross-platform narrative movement. Methodologically, we offer a content-driven, platform-agnostic framework for constructing discourse networks: scalable, unsupervised, and robust to missing behavioral metadata. More broadly, we argue that understanding online discourse requires shifting focus from individual platforms to the connective structures that link them. Ideas flow across ecosystems, and capturing this movement demands tools attuned to the networked nature of modern communication: not just in terms of behavior, but in the deeper discursive alignments that sustain influence. Although prior work has shown that cross-platform contagion is often driven by high-attention or elite accounts~\cite{wilson2020cross, ribeiro2021evolution}, few have examined the structural patterns formed through shared engagement with ideas rather than direct interaction. By representing users as distributions over shared narratives, our framework identifies these latent pathways and the actors who traverse them, linking fragmented publics through structurally embedded roles akin to boundary spanners and brokers. This discourse-centered approach offers a scalable and theory-informed lens to model influence in a fractured information environment.

\vspace{3pt}
\noindent
\textbf{Limitations.}
While our study reveals important patterns in cross-platform narrative migration, it is not without limitations. First, our analysis is limited to two platforms, X and Truth Social, and does not capture the full complexity of the broader information ecosystem. Second, while our semantic clustering approach enables scalable narrative tracking, it may miss highly subtle variations or instances of near-synonymous discursive drift. Finally, although we identify structural convergence and symbolic signaling among key \textit{bridge users}, we cannot definitively infer intentional coordination from observational data alone.




\vspace{3pt}
\noindent
\textbf{Future Work.}
 Future research can extend this framework in several directions. First, incorporating additional platforms such as Telegram, TikTok, Reddit, and Gab would allow for a more comprehensive analysis of cross-ecosystem narrative flow. Second, while this study focuses on user-level discourse, the method is readily adaptable to other entities such as media outlets or organizations, making it possible to trace narrative migration across actors, genres, or communities. Finally, integrating dynamic temporal modeling could uncover how bridge zones form, dissolve, and shift in response to evolving narrative and platform-specific conditions.


\vspace{3pt}
\noindent

\textbf{Ethical Statement.}
This study uses publicly available data from X and Truth Social, collected in accordance with platform terms. Because our analysis examines patterns of influence and discourse across platforms, all usernames are anonymized in the released dataset, and no identifying information is shared. Our focus is on aggregate narrative dynamics and structural features of discourse networks, rather than individual-level behavior. We do not make claims about user intent or coordination. The dataset may contain offensive content given the nature of political discourse, but our analysis focuses on aggregate patterns rather than individual behavior. Data will be released with clear documentation to support open access and reuse for public-interest research on political discourse, narrative diffusion, and cross-platform communication dynamics.

\vspace{3pt}
\noindent
\textbf{Conclusion.} Unlike traditional interaction-based or semantic similarity-based graphs that rely on surface level overlap or direct behavioral ties, CANE models users as distributions over shared narratives. This captures not just what they say or how they interact, but what ideas they inhabit. It reframes network construction from tracing explicit signals to uncovering the conceptual structure that underlies discourse across fragmented ecosystems.

By modeling discourse in this way, we find that cross-platform diffusion between X and Truth Social during the 2024 U.S. election was not only structured, but also predictable: revealing pathways that behavioral and similarity-based methods fail to detect. These patterns point to deeper structures that shape how narratives moved between these fragmented communities. As political discourse continues to fracture, identifying such structures is critical: not just for understanding how information spreads, but for actively anticipating its movement across fragmented publics.

\bibliography{aaai25}
\newpage

\subsection{Paper Checklist to be included in your paper}

\begin{enumerate}

\item For most authors...
\begin{enumerate}
    \item  Would answering this research question advance science without violating social contracts, such as violating privacy norms, perpetuating unfair profiling, exacerbating the socio-economic divide, or implying disrespect to societies or cultures?
     \answerYes{Our study advances computational methods for modeling cross-platform discourse without infringing on user privacy or reinforcing harmful social dynamics. We anonymize all released data and focus on aggregate narrative structures, not individual behavior.}
 
 \item Do your main claims in the abstract and introduction accurately reflect the paper's contributions and scope?
    \answerYes{We state our contributions in narrative tracking, platform-agnostic user graph construction, and identification of bridge users.}
\item Do you clarify how the proposed methodological approach is appropriate for the claims made? 
    \answerYes{Our method is designed to detect cross-platform narrative flow through latent discourse clusters. This aligns  with the claims about discourse structure and user roles across platforms.}
    
\item Do you clarify what are possible artifacts in the data used, given population-specific distributions?
\answerYes{See Limitations: We acknowledge that platform differences and linguistic variability may introduce artifacts; also, we we apply clustering thresholds and human validation to mitigate potential artifacts.}

\item Did you describe the limitations of your work?
\answerYes{Yes, see the Limitations section.}
  \item Did you discuss any potential negative societal impacts of your work?
\answerYes{We caution against misinterpreting structural diffusion patterns as intentional coordination. Additionally, we warn about overreliance on structural roles.}

\item Did you discuss any potential misuse of your work?
\answerYes{See Ethical Statement: We note that identifying bridge users could be misused to target or surveil individuals. We mitigate this risk through full anonymization.}

\item Did you describe steps taken to prevent or mitigate potential negative outcomes of the research, such as data and model documentation, data anonymization, responsible release, access control, and the reproducibility of findings?
\answerYes{We anonymized all data, did not release any user-identifiable information, and documented our methodology in the appendix and supplemental code.}

\item Have you read the ethics review guidelines and ensured that your paper conforms to them?
\answerYes{}
\end{enumerate}

\item Additionally, if your study involves hypotheses testing...
\begin{enumerate}
  \item Did you clearly state the assumptions underlying all theoretical results?
 \answerNA{NA}
 \item Have you provided justifications for all theoretical results?
     \answerNA{NA}
  \item Did you discuss competing hypotheses or theories that might challenge or complement your theoretical results?
    \answerNA{NA}
  \item Have you considered alternative mechanisms or explanations that might account for the same outcomes observed in your study?
    \answerYes{See Bridge Users and Narrative Diffusion: We discuss alternative explanations such as visibility-based influence versus structural positioning.}
  \item Did you address potential biases or limitations in your theoretical framework?
     \answerYes{See Methodology: We acknowledge that structure may not capture all aspects of influence, and that clustering thresholds may miss subtle narrative drift.}
 
 \item Have you related your theoretical results to the existing literature in social science?
     \answerYes{See Related Work, Discussion: We connect our findings to theories of weak ties, boundary spanning, and brokerage.}
  \item Did you discuss the implications of your theoretical results for policy, practice, or further research in the social science domain?
     \answerYes{We discuss implications for incorrect information tracking, platform governance, and cross-platform content moderation.}
\end{enumerate}

\item Additionally, if you are including theoretical proofs...
\begin{enumerate}
  \item Did you state the full set of assumptions of all theoretical results?
    \answerNA{NA}
	\item Did you include complete proofs of all theoretical results?
    \answerNA{NA}
\end{enumerate}

\item Additionally, if you ran machine learning experiments...
\begin{enumerate}
  \item Did you include the code, data, and instructions needed to reproduce the main experimental results (either in the supplemental material or as a URL)?
    \answerYes{We provide anonymized code and datasets through an anonymized GitHub link.}
  \item Did you specify all the training details (e.g., data splits, hyperparameters, how they were chosen)?
    \answerYes{See Appendix.}
     \item Did you report error bars (e.g., with respect to the random seed after running experiments multiple times)?
    \answerYes{We report standard deviations across runs for all evaluations.}
	\item Did you include the total amount of compute and the type of resources used (e.g., type of GPUs, internal cluster, or cloud provider)?
    \answerYes{See Appendix: We include compute details (e.g., use of FAISS with GPU support).}
    
     \item Do you justify how the proposed evaluation is sufficient and appropriate to the claims made? 
    \answerYes{See Experiments: Our evaluation includes multiple downstream tasks across three real-world datasets.}
     \item Do you discuss what is ``the cost`` of misclassification and fault (in)tolerance?
    \answerYes{See Ethical Statement.}
  
\end{enumerate}

\item Additionally, if you are using existing assets (e.g., code, data, models) or curating/releasing new assets, \textbf{without compromising anonymity}...
\begin{enumerate}
  \item If your work uses existing assets, did you cite the creators?
    \answerYes{All datasets and models are appropriately cited.}
  \item Did you mention the license of the assets?
    \answerYes{See Appendix.}
    
  \item Did you include any new assets in the supplemental material or as a URL?
    \answerYes{An anonymized dataset and replication code.}
  \item Did you discuss whether and how consent was obtained from people whose data you're using/curating?
    \answerYes{All data were collected from publicly available sources in accordance with platform terms of service, and  derived from existing datasets released in prior research. We use these datasets with appropriate citations.}

  \item Did you discuss whether the data you are using/curating contains personally identifiable information or offensive content?
    \answerYes{See Ethical Statement: We anonymized all usernames and note that offensive content may be present, but is analyzed only in aggregate.}
    
\item If you are curating or releasing new datasets, did you discuss how you intend to make your datasets FAIR (see \citet{fair})?
 \answerYes{We describe plans for public access, clear documentation, and support for reuse. We include more information in the supplemental material/link.}

\item If you are curating or releasing new datasets, did you create a Datasheet for the Dataset (see \citet{gebru2021datasheets})? 
\answerYes{We include a datasheet in the supplemental material/link.}

\end{enumerate}

\item Additionally, if you used crowdsourcing or conducted research with human subjects, \textbf{without compromising anonymity}...
\begin{enumerate}
  \item Did you include the full text of instructions given to participants and screenshots?
    \answerNA{NA}
  \item Did you describe any potential participant risks, with mentions of Institutional Review Board (IRB) approvals?
    \answerNA{NA}
  \item Did you include the estimated hourly wage paid to participants and the total amount spent on participant compensation?
    \answerNA{NA}
   \item Did you discuss how data is stored, shared, and deidentified?
    \answerNA{NA}
\end{enumerate}

\end{enumerate}
\newpage

\appendix

\section{Appendix A: DP-Means Details}

\paragraph{Why Use DP-Means?}
We adopt DP-Means over alternatives like DBSCAN or agglomerative clustering for its linear time complexity ($O(nKT)$), low memory overhead ($O(K)$), and conceptual clarity. In contrast, DBSCAN has worst-case complexity $O(n^2)$ and often struggles with high-dimensional, sparse embeddings; agglomerative clustering requires $O(n^2 \log n)$ time and is impractical at our scale. Moreover, DP-Means' use of a single interpretable parameter contrasts with the more opaque controls of DBSCAN (e.g., minimum cluster size) and agglomerative methods (e.g., linkage criteria). Finally, DP-Means supports online updates, making it suitable for large-scale or streaming analyses. Full algorithmic details are provided in Appendix A.

\paragraph{DP-Means Objective.}
Given a set of embedded posts $X = \{x_1, ..., x_n\}$ where $x_i \in \mathbf{R}^d$, DP-Means minimizes the following objective:


\begin{equation}
   \min_{z, \mu} \sum_{i=1}^n \min(\min_{k=1}^K d(x_i, \mu_k), \lambda)
\end{equation}

\noindent where $z$ is the cluster assignment, $\mu_k$ are cluster centers, and $d(\cdot,\cdot)$ is cosine distance. A new cluster is created when no existing cluster is within $\lambda$ distance of a point.

\paragraph{Efficiency and Interpretability.}
DP-Means runs in $O(nKT)$ time and $O(K)$ space, supporting online streaming. We favor it over DBSCAN ($O(n^2\log n)$) and agglomerative clustering ($O(n^2d)$) due to its speed, memory efficiency, and direct interpretability via $\lambda$. Unlike opaque hyperparameters (e.g., `minPts' in DBSCAN), $\lambda$ has a direct semantic meaning.

\section{Appendix B: User Network Construction Pipeline Evaluation Details}

\vspace{3pt}
\noindent
\textbf{Weighting Strategy Ablation for CANE.} To understand how different edge weighting schemes affect performance, we evaluate three variants of our CANE graph construction method on the China and Iran information operation datasets: 
(1) \textit{Raw Count}, where edge weights are simple co-participation counts; 
(2) \textit{Softmax-Normalized}, which applies a softmax over per-user neighbors; and 
(3) \textit{TF-IDF-Inspired}, our proposed method which weighs user-cluster participation based on global distinctiveness.

Table~\ref{tab:cane_weighting_ablation} reports the comparative results. Across both datasets, the TF-IDF-inspired weighting strategy consistently outperforms the other two, illutrating the value of incorporating global cluster salience when modeling user similarity.

\begin{table}[t]
    \centering
    \renewcommand{\arraystretch}{1}
    \resizebox{0.48\textwidth}{!}{ 
    
    \begin{tabular}{lcccc}
        \toprule
        \textbf{Weighting Strategy} & \multicolumn{2}{c}{\textbf{China}} & \multicolumn{2}{c}{\textbf{Iran}} \\
        \cmidrule(lr){2-3} \cmidrule(lr){4-5}
        & F1 & AUC & F1 & AUC \\
        \midrule
        Raw Count & 0.70 $\pm$ 0.01 & 0.90 $\pm$ 0.01 & 0.80 $\pm$ 0.01 & 0.93 $\pm$ 0.01 \\
        Softmax-Normalized & 0.54 $\pm$ 0.02 & 0.87 $\pm$ 0.02 & 0.67 $\pm$ 0.01 & 0.86 $\pm$ 0.00 \\
        TF-IDF-Inspired & \textbf{0.72 $\pm$ 0.02} & \textbf{0.91 $\pm$ 0.01} & \textbf{0.82 $\pm$ 0.01} & \textbf{0.94 $\pm$ 0.00} \\
        \bottomrule
    \end{tabular}
    }
    \vspace{2mm}
    \caption{Performance comparison of CANE weighting strategies on the China and Iran information operation datasets. The TF-IDF-inspired scheme outperforms both raw count and softmax-normalized weighting approaches.}
    \label{tab:cane_weighting_ablation}
\end{table}




\vspace{3pt}
\noindent
\textbf{Approximate Nearest Neighbor Evaluation (FAISS).} To scale our CANE graph construction, we replace brute-force similarity comparisons with approximate nearest neighbor search using FAISS~\footnote{\url{https://github.com/facebookresearch/faiss}}. Specifically, we use the Hierarchical Navigable Small World (HNSW) index~\cite{malkov2018efficient}, which enables sublinear retrieval while maintaining high recall. This substitution accelerates user-user similarity computation without compromising downstream performance.

To confirm that approximation does not degrade model quality, we compare the original brute-force CANE construction with its FAISS-based counterpart on the China and Iran IO datasets. As shown in Table~\ref{tab:cane_faiss_eval}, performance remains statistically indistinguishable across both campaigns, validating the use of FAISS-HNSW as a practical and scalable alternative.

\begin{table}[t]
    \centering
    \renewcommand{\arraystretch}{1}
    \resizebox{0.48\textwidth}{!}{
    \begin{tabular}{lcccc}
        \toprule
        \textbf{Similarity Backend} & \multicolumn{2}{c}{\textbf{China}} & \multicolumn{2}{c}{\textbf{Iran}} \\
        \cmidrule(lr){2-3} \cmidrule(lr){4-5}
        & F1 & AUC & F1 & AUC \\
        \midrule
        Brute-force & 0.72 $\pm$ 0.01 & 0.92 $\pm$ 0.02 & 0.83 $\pm$ 0.01 & 0.94 $\pm$ 0.00 \\
        
        FAISS-HNSW & 0.72 $\pm$ 0.02 & 0.91 $\pm$ 0.01 & 0.82 $\pm$ 0.01 & 0.94 $\pm$ 0.00 \\
        \bottomrule
    \end{tabular}
    }
    \vspace{2mm}
    \caption{Comparison of exact (brute-force) vs.\ approximate (FAISS-HNSW) nearest neighbor computation in CANE. FAISS-HNSW yields virtually identical performance while theoretically enabling approximately $\sim$10$\times$ faster construction for large graphs (e.g., 77k users, $k=800$, 48M edges).}
    \label{tab:cane_faiss_eval}
\end{table}

\vspace{3pt}
\noindent
\textbf{Computational and Time Complexity.} We outline the computational complexity of our framework in three key stages:  
(1) post-level embedding, (2) clustering via DP-Means, and (3) user-user network construction.  

All methods, including baselines, require the full embedding of $N$ posts into $d$-dimensional vectors, yielding a shared base cost of $O(Nd)$. Our framework adds a clustering stage, using DP-Means, to form a latent topic structure which improves network expressivity and scalability.

We compare our method against platform-agnostic baselines from previous works that also attempt to connect users based on their embedded content, representing fair baselines for content/platform-agnostic methods.

\begin{itemize}
    \item \textbf{Embedding Averaging}: Each user's messages are embedded, and pairwise user similarity is computed as the average cosine similarity between all messages authored by each user.
    
    \item \textbf{k-NN Embedding Graph}: All messages are embedded and used to construct a k-NN graph; user-user edges are then inferred based on how many of their messages are neighbors.
\end{itemize}

We enumerate this in Table \ref{tab:time-complexity}.

\begin{table}[h]
\centering\renewcommand{\arraystretch}{1.2}
\captionsetup{font=small}
\resizebox{0.47\textwidth}{!}{
\begin{tabular}{lccc}
\toprule
\textbf{Method} & \textbf{Embedding} & \textbf{Clustering} & \textbf{Graph Construction} \\
\midrule
\multicolumn{4}{l}{\textit{Our Methods}} \\\cmidrule(lr){1-4}
CANE (FAISS) & $O(Nd)$ & $O(NKT)$ & $O(U \log U + Uk)$ \\
t-CANE (FAISS) & $O(Nd)$ & $O(NKT)$ & $O(T \cdot (U \log U + Uk))$ \\
\midrule
\multicolumn{4}{l}{\textit{Baselines}} \\\cmidrule(lr){1-4}
Embedding Averaging & $O(Nd)$ & --- & $O(U^2 d)$ \\
k-NN Embedding Graph & $O(Nd)$ & --- & $O(N \log N + kN)$ \\
\bottomrule
\end{tabular}
}
\caption{
Breakdown of computational costs for different network construction methods. 
All methods require embedding $N$ posts into $d$-dimensional space. 
CANE and t-CANE include clustering via DP-Means ($K$ clusters), and use approximate user-user graph construction via FAISS. 
Baseline methods compute similarity at the post or user level without intermediate clustering.
\textbf{Notation:} $N$ - posts, $U$ - users, $d$ - embedding dim, $K$ - clusters, $T$ - time bins, $k$ - neighbors.
}
\label{tab:time-complexity}
\end{table}

\begin{figure}[h]
    \centering
    \resizebox{0.48\textwidth}{!}{
    \begin{tikzpicture}
        \definecolor{CANE_FAISS}{HTML}{afcafc}
        \definecolor{tCANE_FAISS}{HTML}{d3d3d3}
        \definecolor{Embedding_Averaging}{HTML}{0075F2}
        \definecolor{kNN_Embedding_Graph}{HTML}{4CB5AE}

        \begin{semilogyaxis}[ 
            xlabel={Number of Users ($|U|$)},
            ylabel={Relative Computational Cost (log scale)},
            title={Scalability Comparison with Empirical Data},
            ymajorgrids=true,
            grid style={dashed, gray!30},
            tick align=outside,
            legend style={font=\footnotesize, at={(0.5,-0.25)}, anchor=north, cells={align=left}, legend columns=2}
        ]

        \pgfplotstableread[col sep=comma]{csv_data/time_complexity_growth_refined.csv}\datatable

        \addplot[dash dot, thick, CANE_FAISS] table[x=Users, y=CANE_FAISS] {\datatable};
        \addlegendentry{CANE (Ours)}

        \addplot[dash dot, thick, tCANE_FAISS] table[x=Users, y=tCANE_FAISS] {\datatable};
        \addlegendentry{t-CANE (Ours)}

        \addplot[draw, thick, Embedding_Averaging] table[x=Users, y=Embedding_Averaging] {\datatable};
        \addlegendentry{Embedding Averaging (Baseline)}

        \addplot[draw, thick, kNN_Embedding_Graph] table[x=Users, y=kNN_Embedding_Graph] {\datatable};
        \addlegendentry{k-NN Embedding Graph (Baseline)}

        \end{semilogyaxis}
    \end{tikzpicture}
    }
    \captionsetup{font=small}
    \caption{
    Comparison of computational complexity across network construction methods using empirically-informed scaling trends. 
    CANE and t-CANE scale more efficiently with users due to FAISS and clustering, while baselines exhibit steep cost increases tied to post or user volume.
    }
    \label{fig:scalability}
\end{figure}
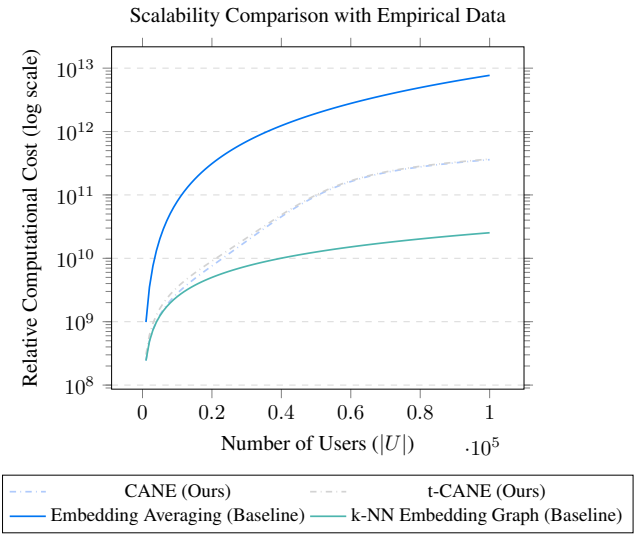

We then use dataset statistics from our experimental settings to empirically model how cluster count, user count, and post volume scale in real-world cases. Based on these trends, we plot the estimated time complexity of each method in Figure~\ref{fig:scalability}. While both of our methods are slightly less efficient than the k-NN Embedding Graph baseline, they are significantly more efficient than the Embedding Averaging approach. As shown in subsequent experiments, we find this minor efficiency tradeoff acceptable given the substantial gains in performance and network expressivity achieved by our methods.

\begin{table}[h]
\centering
\small
\begin{tabular}{lr}
\toprule
\textbf{Threshold} & \textbf{Human Agreement} \\
\midrule
0.60 & 74\%\\
0.65 & 86\%\\
0.70 & 90\%\\
0.75 & 94\%\\
0.80 & 98\%\\
\bottomrule
\end{tabular}
\caption{Human evaluation accuracy for narrative clustering at different cosine similarity thresholds.}
\label{tab:threshold-accuracy-method}
\end{table}

\vspace{3pt}
\noindent
\textbf{Setting $\lambda$: Cosine Similarity Cutoff.} To determine an appropriate cosine similarity threshold for clustering narratives, we conduct a human evaluation of semantic similarity across five thresholds. For each threshold (e.g., 0.60), we sample 50 random paragraph pairs whose cosine similarities fall within a narrow window (e.g., between 0.59 and 0.61). These pairs are annotated by an expert to assess whether they refer to the same underlying topic. The evaluation is conducted exclusively in English, the dominant language in our dataset. Following the protocol outlined in prior work~\cite{hanley2024partial}, we compute human agreement as the proportion of pairs correctly identified as referring to the same or different topics. Table~\ref{tab:threshold-accuracy-method} reports human agreement with model similarity scores across thresholds, providing an empirical basis for selecting a clustering cutoff.

We adopt a threshold of 0.70 as it offers a strong balance between semantic precision and narrative coverage. As shown in Table~\ref{tab:threshold-accuracy-method}, human agreement increases sharply between 0.60 and 0.65, indicating that the model’s similarity scores begin to meaningfully capture topic-level coherence at this point. While stricter thresholds such as 0.75 and 0.80 yield marginally higher agreement, they risk over-segmenting the narrative space by excluding topically related posts that fall just below the cutoff.

\vspace{3pt}
\noindent
\textbf{Clustering Parameter Choice.} To evaluate the sensitivity of our clustering threshold $\lambda$, we conduct a human annotation task across multiple cosine similarity cutoffs. At each threshold, we sample 50 random pairs of posts from different clusters and ask annotators to judge whether they express semantically similar narratives. We report the percentage of agreement with the clustering decision as a proxy for semantic coherence at that threshold. Results in Table~\ref{tab:threshold-accuracy-method} show how human-labeled accuracy varies across similarity cutoffs.

\section{Appendix C: CANE Temporal Extension (t-CANE) and Edge Update Equations}

\paragraph{Edge Update Rule.}
At each time step \( t \), we compute similarity scores \( s_{uv}^t \) from user vectors and update edges as:

\begin{equation}
e_{uv}^t =
\begin{cases}
\alpha \cdot \tilde{s}_{uv}^t + (1 - \alpha) \cdot e_{uv}^{t-1}, & \text{if } s_{uv}^t \text{ is defined} \\
(1 - \beta) \cdot e_{uv}^{t-1}, & \text{otherwise}
\end{cases}
\end{equation}

Here, $\alpha \in [0, 1]$ controls how strongly new similarities influence edge weights, while $\beta \in [0, 1]$ controls the rate at which inactive edges decay. This rewards users who consistently co-engage with similar clusters across timesteps, allowing their edge to strengthen over time, while transient alignments naturally diminish.




\vspace{3pt}
\noindent
\textbf{Hyperparameter Settings.} We set default values of $\alpha = 0.8$ and $\beta = 0.2$, based on downstream validation for information operation detection via grid a targeted grid search. We show example values and corresponding $\text{Macro-F}_1$ and AUC in Table~\ref{tab:china_grid_search}. While these settings generalize well across tasks, we leave more exhaustive hyperparameter optimization to future work.


\vspace{3pt}
\noindent
\textbf{Component Contribution and Ablation.}  
To understand the contribution of each mechanism in \textbf{t-CANE}, we ablate the model’s core components. Specifically, we evaluate performance when removing temporal memory (\(\alpha = 0\)) and decay weighting (\(\beta = 0\)), while keeping other elements fixed. We also compare against simpler aggregation baselines, including unweighted similarity summation and naive averaging over time. 

As shown in Table~\ref{tab:tcane_ablation}, the full model outperforms ablated variants and baselines, underscoring the importance of both memory integration and decay control. Averaging performs reasonably but lacks the capacity to distinguish stable alignment from noise.


\begin{table*}[t]
    \centering
    \small
    \renewcommand{\arraystretch}{1.15}
    \resizebox{0.65\textwidth}{!}{
    \begin{tabular}{lcccc}
        \toprule
        \textbf{Configuration} & \textbf{ROC-AUC} & \textbf{Precision} & \textbf{Recall} & \textbf{$\text{F}_1$} \\
        \midrule
        $\alpha=0.2, \beta=0.0$  & 0.98 $\pm$ 0.00 & 1.00 $\pm$ 0.00 & 0.74 $\pm$ 0.02 & 0.82 $\pm$ 0.02 \\
        $\alpha=0.2, \beta=0.05$ & 0.98 $\pm$ 0.01 & 1.00 $\pm$ 0.00 & 0.74 $\pm$ 0.01 & 0.83 $\pm$ 0.01 \\
        $\alpha=0.2, \beta=0.1$  & 0.98 $\pm$ 0.01 & 1.00 $\pm$ 0.00 & 0.73 $\pm$ 0.02 & 0.81 $\pm$ 0.02 \\
        $\alpha=0.2, \beta=0.15$ & 0.98 $\pm$ 0.01 & 1.00 $\pm$ 0.00 & 0.73 $\pm$ 0.01 & 0.81 $\pm$ 0.01 \\
        $\alpha=0.2, \beta=0.2$  & \textbf{0.98} $\pm$ \textbf{0.01} & 1.00 $\pm$ 0.00 & 0.74 $\pm$ 0.02 & 0.82 $\pm$ 0.02 \\
        $\alpha=0.3, \beta=0.0$  & 0.98 $\pm$ 0.01 & 1.00 $\pm$ 0.00 & 0.74 $\pm$ 0.01 & 0.82 $\pm$ 0.01 \\
        $\alpha=0.3, \beta=0.05$ & 0.98 $\pm$ 0.01 & 1.00 $\pm$ 0.00 & 0.74 $\pm$ 0.02 & 0.82 $\pm$ 0.02 \\
        $\alpha=0.3, \beta=0.1$  & \textbf{0.98} $\pm$ \textbf{0.01} & 1.00 $\pm$ 0.00 & 0.74 $\pm$ 0.01 & 0.82 $\pm$ 0.01 \\
        $\alpha=0.3, \beta=0.15$ & 0.98 $\pm$ 0.01 & 1.00 $\pm$ 0.00 & 0.74 $\pm$ 0.01 & 0.82 $\pm$ 0.01 \\
        $\alpha=0.3, \beta=0.2$  & 0.98 $\pm$ 0.00 & 1.00 $\pm$ 0.00 & 0.74 $\pm$ 0.01 & 0.82 $\pm$ 0.01 \\
        $\alpha=0.4, \beta=0.0$  & 0.98 $\pm$ 0.01 & 1.00 $\pm$ 0.00 & 0.74 $\pm$ 0.01 & 0.82 $\pm$ 0.01 \\
        $\alpha=0.4, \beta=0.05$ & 0.98 $\pm$ 0.01 & 1.00 $\pm$ 0.00 & 0.74 $\pm$ 0.01 & 0.83 $\pm$ 0.01 \\
        $\alpha=0.4, \beta=0.1$  & 0.98 $\pm$ 0.01 & 1.00 $\pm$ 0.00 & 0.73 $\pm$ 0.02 & 0.82 $\pm$ 0.02 \\
        $\alpha=0.4, \beta=0.15$ & 0.98 $\pm$ 0.01 & 1.00 $\pm$ 0.00 & 0.74 $\pm$ 0.01 & 0.82 $\pm$ 0.01 \\
        $\alpha=0.4, \beta=0.2$  & 0.98 $\pm$ 0.01 & 1.00 $\pm$ 0.00 & 0.74 $\pm$ 0.02 & 0.83 $\pm$ 0.01 \\
        \midrule
        $\alpha=0.5, \beta=0.05$ & 0.98 $\pm$ 0.01 & 1.00 $\pm$ 0.00 & 0.73 $\pm$ 0.02 & 0.82 $\pm$ 0.02 \\
        $\alpha=0.5, \beta=0.1$  & 0.98 $\pm$ 0.01 & 1.00 $\pm$ 0.00 & 0.73 $\pm$ 0.01 & 0.81 $\pm$ 0.01 \\
        $\alpha=0.6, \beta=0.05$ & 0.98 $\pm$ 0.01 & 1.00 $\pm$ 0.00 & 0.73 $\pm$ 0.02 & 0.81 $\pm$ 0.01 \\
        $\alpha=0.6, \beta=0.1$  & 0.98 $\pm$ 0.01 & 1.00 $\pm$ 0.00 & 0.74 $\pm$ 0.01 & 0.82 $\pm$ 0.01 \\
        $\alpha=0.7, \beta=0.05$ & 0.98 $\pm$ 0.01 & 1.00 $\pm$ 0.00 & 0.74 $\pm$ 0.02 & 0.82 $\pm$ 0.02 \\
        $\alpha=0.7, \beta=0.1$  & 0.97 $\pm$ 0.01 & 1.00 $\pm$ 0.00 & 0.73 $\pm$ 0.02 & 0.81 $\pm$ 0.02 \\
        $\alpha=0.7, \beta=0.2$  & 0.98 $\pm$ 0.01 & 1.00 $\pm$ 0.00 & 0.74 $\pm$ 0.02 & 0.83 $\pm$ 0.02 \\
        $\alpha=0.8, \beta=0.05$ & 0.98 $\pm$ 0.01 & 1.00 $\pm$ 0.00 & 0.73 $\pm$ 0.02 & 0.82 $\pm$ 0.02 \\
        $\alpha=0.8, \beta=0.1$  & 0.98 $\pm$ 0.01 & 1.00 $\pm$ 0.00 & 0.74 $\pm$ 0.02 & 0.82 $\pm$ 0.02 \\
        $\alpha=0.8, \beta=0.15$ & 0.98 $\pm$ 0.00 & 1.00 $\pm$ 0.00 & 0.73 $\pm$ 0.02 & 0.82 $\pm$ 0.01 \\
        $\alpha=0.8, \beta=0.2$  & \textbf{0.98} $\pm$ \textbf{0.01} & \textbf{1.00} $\pm$ \textbf{0.00} & \textbf{0.75} $\pm$ \textbf{0.01} & \textbf{0.83} $\pm$ \textbf{0.01} \\
        \bottomrule
    \end{tabular}
    }
    \caption{Expanded grid search results on the China Information Operations detection task. All metrics are macro-averaged and reported as mean $\pm$ standard deviation across cross-validation folds.}
    \label{tab:china_grid_search}
\end{table*}

\begin{table*}[h]
    \centering
    \renewcommand{\arraystretch}{1.15}
    \resizebox{0.85\textwidth}{!}{
    \begin{tabular}{lcccc}
        \toprule
        \textbf{Model Variant} & \textbf{ROC-AUC} & \textbf{Precision} & \textbf{Recall} & \textbf{$\text{F}_1$} \\
        \midrule
        Full t-CANE & \textbf{0.98 $\pm$ 0.01} & \textbf{1.00 $\pm$ 0.00} & \textbf{0.75 $\pm$ 0.01} & \textbf{0.83 $\pm$ 0.01} \\
        \midrule
        \textit{w/o Memory} ($\alpha{=}0$) & 0.00 $\pm$ 0.00 & 0.00 $\pm$ 0.00 & 0.00 $\pm$ 0.00 & 0.00 $\pm$ 0.00 \\
        \textit{w/o Decay} ($\beta{=}0$) & 0.98 $\pm$ 0.01 & 1.00 $\pm$ 0.00 & 0.73 $\pm$ 0.02 & 0.81 $\pm$ 0.02 \\
        \textit{No Memory or Decay} (Average Similarity) & 0.96 $\pm$ 0.01 & 1.00 $\pm$ 0.00 & 0.72 $\pm$ 0.02 & 0.79 $\pm$ 0.02 \\
        \textit{No Memory or Decay} (Naive Aggregation) & 0.97 $\pm$ 0.00 & 0.99 $\pm$ 0.00 & 0.71 $\pm$ 0.02 & 0.80 $\pm$ 0.01 \\
        \bottomrule
    \end{tabular}
    }
    \caption{Ablation study on key components of \textbf{t-CANE} based on results from the China Information Operations detection task. All metrics are macro-averaged.
The full model incorporates temporal memory ($\alpha$) and decay smoothing ($\beta$). Removing memory ($\alpha{=}0$) leads to a collapse in performance, as user similarity cannot meaningfully propagate across time. The final two rows reflect ablations of both components, where user similarity is computed independently at each timestep and aggregated via averaging or summation.
}
    \label{tab:tcane_ablation}
\end{table*}



\section{Appendix D: Information Operations Results}
\label{sec:appendix-io}

\vspace{1mm}
\noindent
\textbf{Dataset Composition.} We evaluate on Twitter’s publicly released Information Operations archive~\cite{seckin2024labeled}, using labeled campaigns attributed to China and Iran. Each dataset includes state-backed IO driver accounts and a large set of matched organic control users. Table~\ref{tab:io_users} summarizes the number of accounts and total posts per group.

\begin{table}[h]
\small
\centering
\renewcommand{\arraystretch}{1.1}

\begin{tabular}{lll}
\toprule
\textbf{Country} & \textbf{IO Drivers [Posts]} & \textbf{Control Users [Posts]} \\
\midrule
China & 5,191 [13.8M] & 76,286 [3.5M] \\
Iran  & 209 [9.9M] & 16,885 [2.5M] \\
\bottomrule
\end{tabular}
\vspace{1mm}
\caption{IO datasets used in this study. For each campaign, we report the number of state-backed accounts (IO drivers) and control users, along with post volumes.}
\label{tab:io_users}
\end{table}

\vspace{2mm}
\noindent
\textbf{Network Construction Details.} Following~\citet{luceri2024unmasking}, we generate user-user similarity graphs for each network construction method. For our methods, CANE and t-CANE, we compute participation profiles over clustered topics and connect users based on cosine similarity. Temporal variants further aggregate user behavior over time windows.

\vspace{2mm}
\noindent
\textbf{Evaluation Setup.} Each graph is embedded using node2vec~\cite{grover2016node2vec}, and a random forest classifier is trained to predict whether a user is part of the IO campaign. Models are evaluated using 5-fold cross-validation, with Macro-F$_1$ and AUC reported. Full results are shown in Tables~\ref{tab:china_io} and~\ref{tab:iran_io}.

\vspace{2mm}
\noindent
\textbf{Sparse Data Robustness.} To evaluate data efficiency, we simulate limited post availability by sampling $n\%$ of each user's content in 5\% increments. At each step, we measure when a method reaches 95\% of its best AUC. Table~\ref{tab:data_efficiency_95auc_IO} reports these thresholds, demonstrating that CANE t-CANE achieves strong performance with minimal data.

\begin{table}[h]
\centering
\renewcommand{\arraystretch}{1.1}
\resizebox{0.4\textwidth}{!}{
\begin{tabular}{llcc}
\toprule
\textbf{Category} & \textbf{Method} & \textbf{China} & \textbf{Iran} \\
\midrule
\multirow{7}{*}{Baseline} 
& Co-Repost & 15\% & 20\% \\
& Co-URL & 20\% & 25\% \\
& Fast Repost & 20\% & 20\% \\
& Hashtag Sequence & 80\% & 60\% \\
& Text Similarity & \textbf{5\%} & 10\% \\
& k-NN Embedding Graph & 10\% & \textbf{5\%} \\
& Fused Graph & 15\% & 30\% \\
\midrule
\multirow{2}{*}{Ours} 
& CANE & \textbf{5\%} & 10\% \\
& t-CANE & \textbf{5\%} & \textbf{5\%} \\
\bottomrule
\end{tabular}
}
\vspace{1mm}
\caption{Percentage of training data required to reach 95\% of peak AUC for each method on the China and Iran IO campaigns. Lower values reflect higher data efficiency.}
\label{tab:data_efficiency_95auc_IO}
\end{table}

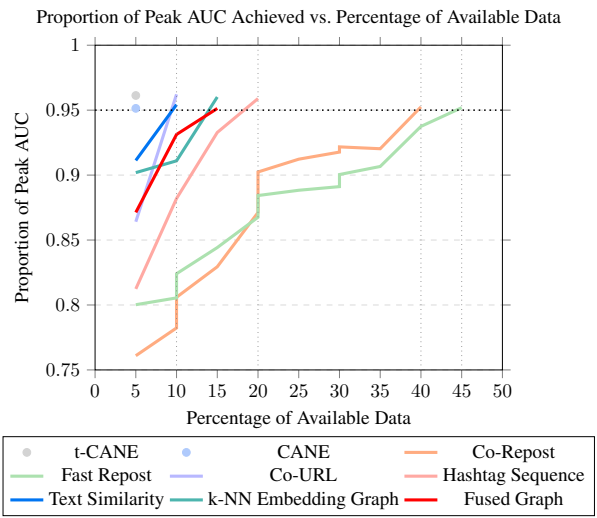
\begin{figure}[h]
\resizebox{0.45\textwidth}{!}{
\centering
\begin{tikzpicture}
\begin{axis}[
    xlabel={Percentage of Available Data},
    ylabel={Proportion of Peak AUC},
    title={Proportion of Peak AUC Achieved vs. Percentage of Available Data},
    ymajorgrids=true,
    grid style={dashed, gray!30},
    xtick={0,5,10,15,20,25,30,35,40,45,50},
    tick align=outside,
    legend style={at={(0.5,-0.45)}, anchor=south, cells={align=left}, legend columns=3},
    xmin=0,
    ymin=0.75,
    xmax=50,
    ymax=1,
    width=9cm,
    height=7.5cm,
]

        \definecolor{co-url}{HTML}{b8b8ff}
        \definecolor{co-hashtag}{HTML}{ffa9a3}
        \definecolor{co-retweet}{HTML}{ffac81}
        \definecolor{co-fast-retweet}{HTML}{ace1af}
        \definecolor{CANE_FAISS}{HTML}{afcafc}
        \definecolor{tCANE_FAISS}{HTML}{d3d3d3}
        \definecolor{Embedding_Averaging}{HTML}{0075F2}
        \definecolor{kNN_Embedding_Graph}{HTML}{4CB5AE}

\addplot+[
    color=tCANE_FAISS,
    fill=tCANE_FAISS,
    mark=*,
    only marks,
    mark options={fill=tCANE_FAISS, draw=tCANE_FAISS},
    mark size=2pt
] coordinates {
    (5, 0.961289)
};

\addlegendentry{t-CANE}

\addplot+[
    color=CANE_FAISS,
    fill=CANE_FAISS,
    mark=*,
    only marks,
    mark options={fill=CANE_FAISS, draw=CANE_FAISS},
    mark size=2pt
] coordinates {
    (5, 0.95138)
};
\addlegendentry{CANE}

\addplot+[solid, color=co-retweet, ultra thick, mark=none] table[
    x=percent,
    y=distance_from_peak,
    col sep=comma
] {csv_data/deg_co-retweet-china.csv};
\addlegendentry{Co-Repost}

\addplot+[solid, color=co-fast-retweet, ultra thick, mark=none] table[
    x=percent,
    y=distance_from_peak,
    col sep=comma
] {csv_data/deg_co-fast-retweet-china.csv};
\addlegendentry{Fast Repost}

\addplot+[solid, color=co-url, ultra thick, mark=none] table[
    x=percent,
    y=distance_from_peak,
    col sep=comma
] {csv_data/deg_co-url-china.csv};
\addlegendentry{Co-URL}

\addplot+[solid, color=co-hashtag, ultra thick, mark=none] table[
    x=percent,
    y=distance_from_peak,
    col sep=comma
] {csv_data/deg_co-hashtag-china.csv};
\addlegendentry{Hashtag Sequence}

\addplot+[solid, color=Embedding_Averaging, ultra thick, mark=none] table[
    x=percent,
    y=distance_from_peak,
    col sep=comma
] {csv_data/deg-text-sim-china.csv};
\addlegendentry{Text Similarity}

\addplot+[solid, color=kNN_Embedding_Graph, ultra thick, mark=none] table[
    x=percent,
    y=distance_from_peak,
    col sep=comma
] {csv_data/deg-knn-china.csv};
\addlegendentry{k-NN Embedding Graph}

\addplot+[solid, color=red, ultra thick, mark=none] table[
    x=percent,
    y=distance_from_peak,
    col sep=comma
] {csv_data/deg-fused-graph-china.csv};
\addlegendentry{Fused Graph}

\addplot[dotted, black, thick] coordinates {(0,0.95) (100,0.95)};

\addplot[dotted, gray] coordinates {(10,1) (10,0.7)};

\addplot[dotted, gray] coordinates {(20,1) (20,0.7)};
\addplot[dotted, gray] coordinates {(40,1) (40,0.7)};
\addplot[dotted, gray] coordinates {(45,1) (45,0.7)};

\end{axis}
\end{tikzpicture}
}
\caption{Proportion of peak AUC achieved as a function of training data percentage. Each curve is truncated at the first point where the method reaches 95\% or more of its peak performance, illustrating the data required to achieve near-optimal results. }
\label{fig:peak-auc}
\end{figure}

\vspace{2mm}
\noindent
\textbf{Interpreting Peak AUC Thresholds.} Figure~\ref{fig:peak-auc} visualizes the proportion of peak AUC achieved as a function of available training data. Each method's curve is truncated at the earliest point where 95\% of its maximum AUC is reached. This threshold balances interpretability and robustness: the trendlines shows smooth monotonic increases, and most methods saturate well before full data availability. For instance, both CANE and t-CANE reach 95\% of peak performance with just 5\% of user data. These saturation points validate the 95\% threshold as a conservative but effective marker of data efficiency.

\section{Appendix E: Annotation Statistics for Ideological Mapping}
\label{sec:ideology-classification-annotations}

\vspace{3pt}
\noindent
\textbf{Initial Annotation and Agreement.} Table~\ref{tab:annotation-stats-twitter-tiktok} presents the initial ideological annotations of users from X and TikTok. Each user was labeled as \textit{Liberal}, \textit{Conservative}, or \textit{Other/NA}.Inter annotator agreement, measured by Cohen’s $\kappa$, was high across both platforms, with values of 0.84 for X and 0.87 for TikTok, indicating strong reliability.

\begin{table}[h]
    \centering
    \renewcommand{\arraystretch}{1}
    \resizebox{0.45\textwidth}{!}{
    \begin{tabular}{lllll}
        \toprule
        \textbf{Label} & \multicolumn{2}{c}{\textbf{X}} & \multicolumn{2}{c}{\textbf{TikTok}} \\
        \cmidrule(lr){2-3} \cmidrule(lr){4-5}
        & Count (\%) & Cohen’s $\kappa$ & Count (\%) & Cohen’s $\kappa$ \\
        \midrule
        Liberal & 503 (24\%) & 0.86 & 408 (23\%) & 0.88 \\
        Conservative & 1,161 (55\%) & 0.88 & 681 (39\%) & 0.89 \\
        Other / NA & 436 (21\%) & 0.79 & 670 (38\%) & 0.85 \\
        \midrule
        \textbf{Total / Avg} & \textbf{2,100} & \textbf{0.84} & \textbf{1,759} & \textbf{0.87} \\
        \bottomrule
    \end{tabular}}
    \caption{Initial annotation statistics for X and TikTok users. Counts and inter-annotator agreement are shown by label.}
    \label{tab:annotation-stats-twitter-tiktok}
\end{table}

\vspace{3pt}
\noindent
\textbf{Filtered High-Confidence Labels.} For downstream modeling, we retain only users where both annotators agreed. Table~\ref{tab:final-label-distribution} shows the final label distributions after filtering. This filtered set serves as the basis for all supervised experiments and classification tasks.

\begin{table}[h]
    \centering
    \renewcommand{\arraystretch}{1}
    \resizebox{0.4\textwidth}{!}{
    \begin{tabular}{lllll}
        \toprule
        \textbf{Label} & \multicolumn{2}{c}{\textbf{X}} & \multicolumn{2}{c}{\textbf{TikTok}} \\
        \cmidrule(lr){2-3} \cmidrule(lr){4-5}
                      & Count & \%       & Count & \%       \\
        \midrule
        Liberal        & 433   & 22.8\%   & 359   & 22.2\%   \\
        Conservative   & 1,057 & 55.6\%   & 620   & 38.4\%   \\
        Other / NA     & 410   & 21.6\%   & 636   & 39.4\%   \\
        \midrule
        \textbf{Total} & \textbf{1,900} & 100\% & \textbf{1,615} & 100\% \\
        \bottomrule
    \end{tabular}}
    \caption{Final label distributions after removing all users with annotator disagreement. Only users with consensus labels are retained.}
    \label{tab:final-label-distribution}
\end{table}

\vspace{3pt}
\noindent
\textbf{Controlled Evaluation for Network Comparison.} In Section~\ref{sec:ideology-classification} \textit{Ideological Mapping on X and TikTok}, we compare network construction methods by training classifiers to predict user ideology from node embeddings. However, raw comparisons are complicated by differing graph coverage (e.g., co-URL and co-Repost) graphs include only small subsets of users due to sparse signals on X.

To isolate model performance from coverage effects, we conduct a controlled evaluation using only the users included in each graph’s subgraph. Table~\ref{tab:annotation-stats-controlled} reports the number of Liberal, Conservative, and Other/NA users in each graph’s subset. For classification experiments, we exclude Other/NA users and balance the liberal and conservative classes to prevent skewed results.

\begin{table}[t]
\centering
\renewcommand{\arraystretch}{1}
\resizebox{0.47\textwidth}{!}{
\begin{tabular}{lccc}
    \toprule
    \textbf{Similarity Network} & \textbf{Liberal} & \textbf{Conservative} & \textbf{Other / NA} \\
    \midrule
    Co-Repost         & 182 & 428 & 220 \\
    Co-URL            & 160 & 269 & 109 \\
    Fast Repost       & 270 & 161 & 72 \\
    Hashtag Sequence  & 331 & 421 & 108 \\
    Fused Graph       & 681 & 483 & 217 \\
    \bottomrule
\end{tabular}
}
\caption{Label distribution in each similarity network’s test subset. Controlled evaluation (Section~\S\ref{sec:ideology-classification}) uses only users with Liberal or Conservative labels in each subset.}
\label{tab:annotation-stats-controlled}
\end{table}

\section{Appendix F: Ideological Mapping Results}
\label{sec:testing-procedure-ideological-classification}

\vspace{3pt}
\noindent
\textbf{Testing Procedure for Ideological Mapping.} We frame the ideology classification task as a purely network-based prediction problem, using only the structure of user-user graphs to predict ideological leaning. While recent studies often combine network and content-based signals~\cite{jiang2023retweet}, we deliberately exclude all text features to isolate the structural contribution of different graph construction methods.

We evaluate our approach on annotated data from X and TikTok, as shown in Section~\ref{sec:ideology-classification}. The underlying label distributions are summarized in Appendix E~\ref{sec:ideology-classification-annotations} (Tables~\ref{tab:annotation-stats-twitter-tiktok} through ~\ref{tab:final-label-distribution}). For model training, we focus only on users labeled as \textit{Liberal} or \textit{Conservative}, discarding all \textit{Other/NA} labels. To ensure that results are not driven by class imbalance, we undersample the majority class within each graph subset to achieve balanced training sets.

We compare our method against several traditional and hybrid network construction baselines: Co-Repost, Co-URL, Fast Repost, Hashtag Sequence, Text Similarity, the k-NN Embedding Graph, and a Fused Graph combining available signals. Due to metadata constraints, the TikTok dataset includes only Hashtag Sequence, Text Similarity (from video descriptions), k-NN Embedding Graph, and a partial Fused Graph constructed from these sources.

To evaluate classification performance, we train three types of graph-based models on each network:
(1) Graph Attention Networks (GAT)~\cite{velivckovic2017graph}, (2) Graph Convolutional Networks (GCN)~\cite{zhang2019graph}, and (3) node2vec~\cite{grover2016node2vec} embeddings combined with a feedforward neural network (NN) classifier. We initially tested node2vec with a random forest (RF) classifier as well, but found it consistently underperformed and therefore excluded it from subsequent comparisons. Overall, performance across GAT, GCN, and node2vec+NN was broadly comparable; for consistency and ease of interpretation, we report main results using node2vec+NN.

All models are trained using 5-fold cross-validation and evaluated on held-out validation sets using $\text{Macro-F}_1$ and AUC. Statistical comparisons between models are conducted using unpaired t-tests on the fold-level results. Performance results (mean and standard deviation) are reported in Table~\ref{tab:similarity_networks-twitter-political}. Full model hyperparameters and training code are available in our public repository.

\vspace{3pt}
\noindent
\textbf{Annotation Guidelines.} Table~\ref{tab:coding-guidelines} reproduces the full labeling instructions shown to annotators for the ideology classification task. These were presented in advance of annotation and were standardized across annotators.

\begin{table}[h]
\centering
\caption{Annotation Instructions Presented to Annotators}
\label{tab:coding-guidelines}
\begin{tabular}{p{0.95\linewidth}}
\toprule
\textbf{Guidelines} \\
\midrule
You will be assigned a list of 150 users. For each user, you will be shown a sample of five posts. Based on these posts, assign one of the following ideological labels to the user: \textit{Liberal}, \textit{Conservative}, or \textit{Other/NA}.

These posts are drawn from a dataset of political discourse related to the 2024 U.S. Presidential Election on X (formerly Twitter). In many cases, a user’s political leaning will be evident through partisan language, endorsements, or other clear cues. However, if a user’s stance is unclear, ambiguous, or unrelated to politics, label them as \textit{Other/NA}.

We acknowledge that this labeling framework is inherently limited. The liberal/conservative binary does not capture political ideology's full complexity. Given the constraints of the task, however, please do your best to assign the most appropriate label based on available evidence. \\
\bottomrule
\end{tabular}
\end{table}

\vspace{3pt}
\noindent
\textbf{Controlled Evaluation Across Network Types.} To isolate the effect of graph structure from node coverage, we rerun our ideological classification experiments using only the intersection of users shared across all graph types. This controlled setup allows us to evaluate structural quality independent of user inclusion, ensuring a fair comparison. As shown in Table~\ref{tab:method-vs-network}, our method performs on par with or better than established graphs, validating that performance gains are not simply a result of broader user coverage.

\begin{table}[h]
    \centering
    \renewcommand{\arraystretch}{1}
    \resizebox{0.47\textwidth}{!}{
    \begin{tabular}{llcc}
        \toprule
        \textbf{Method} & \textbf{Similarity Network} & \textbf{F1} & \textbf{AUC} \\
        \midrule
        Baseline & Co-Repost & 0.82 $\pm$ 0.02 & 0.68 $\pm$ 0.05 \\
        CANE & Co-Repost & 0.79 $\pm$ 0.04 & 0.70 $\pm$ 0.09 \\
        t-CANE & Co-Repost & \textbf{0.86 $\pm$ 0.02} & \textbf{0.83 $\pm$ 0.03} \\
        \cmidrule(lr){1-4}
        Baseline & Co-URL & 0.77 $\pm$ 0.01 & 0.54 $\pm$ 0.03 \\
        CANE & Co-URL & 0.78 $\pm$ 0.04 & 0.72 $\pm$ 0.05 \\
        t-CANE & Co-URL & \textbf{0.82 $\pm$ 0.02} & \textbf{0.79 $\pm$ 0.05} \\
        \cmidrule(lr){1-4}
        Baseline & Fast Repost & 0.88 $\pm$ 0.02 & 0.81 $\pm$ 0.05 \\
        CANE & Fast Repost & 0.83 $\pm$ 0.02 & 0.77 $\pm$ 0.04 \\
        t-CANE & Fast Repost & \textbf{0.89 $\pm$ 0.03} & \textbf{0.87 $\pm$ 0.02} \\
        \cmidrule(lr){1-4}
        Baseline & Hashtag Sequence & 0.81 $\pm$ 0.02 & 0.83 $\pm$ 0.03 \\
        CANE & Hashtag Sequence & 0.77 $\pm$ 0.04 & 0.78 $\pm$ 0.05 \\
        t-CANE & Hashtag Sequence & \textbf{0.83 $\pm$ 0.02} & \textbf{0.84 $\pm$ 0.03} \\
        \cmidrule(lr){1-4}
        Baseline & Fused Graph &\textbf{0.86 $\pm$ 0.01} & \textbf{0.84 $\pm$ 0.02} \\
        CANE & Fused Graph & 0.77 $\pm$ 0.04 & 0.78 $\pm$ 0.05 \\
        t-CANE & Fused Graph & 0.83 $\pm$ 0.02 & \textbf{0.84 $\pm$ 0.03} \\
        \bottomrule
    \end{tabular}
    }
    \caption{Controlled evaluation of ideological classification performance on X, using only the subset of users present in each similarity network’s graph. This setup isolates structural quality by holding user coverage constant across methods. Under these conditions, our method performs comparably to established approaches like Repost and Fused Graphs, indicating that it offers similar structural effectiveness while providing broader user coverage. We omit Text Similarity and k-NN Embedding Graph methods from this comparison, as they naturally include the same users as our approach and require no such control.}
    \label{tab:method-vs-network}
    
\end{table}

\vspace{2mm}
\noindent
\textbf{Data Efficiency Analysis.} To evaluate how robust each method is under limited data conditions, we progressively subsample user posts and measure the minimum percentage of data needed to reach 95\% of each method’s peak AUC. Table~\ref{tab:data_efficiency_95auc_IO-ideological} summarizes these thresholds for both X and TikTok. Our methods consistently achieve near-optimal performance with substantially less data, demonstrating strong data efficiency and robustness to real-world sparsity.

\begin{table}[h]
    \centering
    \renewcommand{\arraystretch}{1}
    \resizebox{0.4\textwidth}{!}{ 
    \begin{tabular}{llcc}
        \toprule
        \textbf{Category} & \textbf{Method} & \textbf{X} & \textbf{TikTok} \\
        \midrule
        \multirow{6}{*}{Baseline} 
        & Co-Repost & 15\% & -- \\
        & Co-URL & 20\% & -- \\
        & Fast Repost & 20\% & -- \\
        & Hashtag Sequence & 80\% & 60\% \\
        & Text Similarity & \textbf{5\%} & 10\% \\
        & k-NN Embedding Graph &  10\% & \textbf{5\%} \\
        & Fused Graph & 15\% & 30\% \\
        \midrule
        \multirow{2}{*}{Ours} 
        & CANE & \textbf{5\%} & 10\% \\
        & t-CANE & \textbf{5\%} & \textbf{5\%} \\
        \bottomrule
    \end{tabular}
    }
    \vspace{2mm}
    \caption{Percentage of training data required to reach 95\% of peak AUC for each method on the X and TikTok datasets. Lower values indicate faster convergence toward near-optimal performance.}
    \label{tab:data_efficiency_95auc_IO-ideological}
\end{table}

\section{Appendix G: Cross-Platform Engagement Prediction Task}

\vspace{2mm}
\noindent
\textbf{Narrative Identification.}  
Narratives are identified via multilingual MPNet embeddings fine-tuned for semantic similarity, followed by DP-means clustering with a cosine distance threshold of 0.35. This threshold was validated through manual evaluation for coherence and specificity. Each cluster was assigned a topic label via top TF-IDF terms. To prevent circularity, narrative text was excluded from graph features.

\vspace{2mm}
\noindent
\textbf{Dataset Summary.} Table~\ref{tab:narrative-summary} provides summary statistics for the full set of narrative themes evaluated in the engagement prediction task. We report the number of users and posts associated with each theme, as well as platform composition and distributional metrics. This context helps characterize the underlying structure and balance of the prediction space across X and Truth Social.

\vspace{2mm}
\noindent
\textbf{Representative Narrative Themes.} Table~\ref{tab:example-narratives} presents a sample of narrative themes with substantial activity across both platforms. These examples illustrate the mix of episodic and persistent content our model handles: ranging from emergent protest events to recurring partisan issues. Platform distribution is also shown to highlight asymmetries in narrative engagement.

\vspace{2mm}
\noindent
\textbf{Model Architecture and Training Setup.} Table~\ref{tab:engagement_gcn} outlines the architecture and training parameters used in our engagement prediction model. We adopt a two-layer Graph Convolutional Network (GCN) with binary input features capturing user-topic interaction histories. Evaluation relies on Micro-F$_1$, AUC, and accuracy, with thresholds selected via validation sweep. A randomly rewired GCN serves as a structural baseline.

We use a two-layer Graph Convolutional Network (GCN)~\cite{zhang2019graph} to predict future engagement with trending topics. Each user is represented by a binary feature vector indicating past engagement with a consistent set of clusters, and edges reflect user similarity based on the selected network (e.g., Co-URL, Fused Graph).

We initially tested multiple architectures, including Graph Attention Networks (GAT), but observed no meaningful performance gains over GCN. For clarity and reproducibility, we report final results using GCN.

We train the model using binary cross-entropy loss with dynamic class weighting to address label imbalance, optimizing over 100 epochs using the Adam optimizer. Evaluation is conducted every 5 epochs, with micro-F$_1$ optimized via threshold sweeping on the validation set. To contextualize performance, we include, as a baseline, a randomly initialized GCN trained on a degree-preserving rewiring of the graph. Performance metrics (F$_1$, AUC, and accuracy) are reported as the mean and standard deviation across snapshots.

\vspace{2mm}
\noindent
\textbf{Engagement Prediction Across Time Windows.} Table~\ref{tab:full-engagement-prediction} presents the performance of different similarity networks in predicting future user engagement with specific narratives across four time horizons: 3, 5, 7, and 14 days. Performance is measured using F$_1$ and AUC.

\begin{table}[t]
\centering
\small
\begin{tabular}{lr}
\toprule
\textbf{Metric} & \textbf{Value} \\
\midrule
Total Narrative Themes & 321 \\
Total Posts & 374{,}148 \\
Total Users & 261{,}398 \\
Median Posts per Theme & 154 \\
Median Users per Theme & 116 \\
Mean X Post \% & 72.3 \\
Mean Truth Social Post \% & 27.7 \\
Median X User Count & 89 \\
Median Truth Social User Count & 15 \\
\bottomrule
\end{tabular}
\caption{Summary statistics for all tested narrative themes.}
\label{tab:narrative-summary}
\end{table}

\begin{table*}[t]
\centering
\small
\begin{tabular}{lrrrrrr}
\toprule
\textbf{Narrative Theme} & \textbf{Posts} & \textbf{Users} & \textbf{X (\%)} & \textbf{TS (\%)} & \textbf{X Users} & \textbf{TS Users} \\
\midrule
Columbia University and Protest Suppression & 688 & 382 & 60.9 & 39.1 & 325 & 57 \\
Promotion of Conservative Media & 518 & 271 & 50.0 & 50.0 & 243 & 29 \\
FBI Investigations and Federal Overreach & 302 & 199 & 57.0 & 43.0 & 162 & 37 \\
Chemical Leaks and Environmental Disasters & 289 & 154 & 55.7 & 44.3 & 125 & 29 \\
Condemnation of Pro-Palestinian Protest Tactics & 329 & 140 & 50.5 & 49.5 & 102 & 39 \\
Mar-a-Lago Probe and GOP Investigations & 247 & 159 & 56.3 & 43.7 & 123 & 38 \\
Domestic Shootings and Public Safety & 239 & 118 & 43.1 & 56.9 & 85 & 33 \\
Nuclear Energy Conspiracy Theories & 89 & 73 & 61.8 & 38.2 & 51 & 22 \\
Criticism of Jan 6 Committee Chair & 87 & 68 & 60.9 & 39.1 & 52 & 18 \\
Global Unrest and War Escalations & 81 & 57 & 49.4 & 50.6 & 36 & 21 \\
Comparing Jan 6 and Tiananmen Square & 66 & 47 & 57.6 & 42.4 & 33 & 14 \\
Transgender Policy Debate & 53 & 42 & 52.8 & 47.2 & 27 & 15 \\
\bottomrule
\end{tabular}
\caption{Examples of narrative themes with presence on both X and Truth Social (TS), shown to illustrate our inclusion of both emerging events and broader, persistent discourse. Narrative themes were manually assigned by a human annotator based on the five posts closest to each narrative cluster centroid and the top TF-IDF terms. Full metadata is available in the accompanying repository.}
\label{tab:example-narratives}
\end{table*}

\vspace{1em}
\begin{table}[h]
\centering
\small
\renewcommand{\arraystretch}{1.1}
\resizebox{0.45\textwidth}{!}{ 
\begin{tabular}{ll}
\toprule
\textbf{Setting} & \textbf{Value} \\
\midrule
Model Type & 2-layer Graph Convolutional Network (GCN) \\
Hidden Dimension & 64 \\
Input Features & Prior engagement (binary per topic) \\
Output & Future engagement (binary per topic) \\
Loss Function & Binary Cross-Entropy with class weights \\
Optimizer & Adam \\
Learning Rate & 0.01 \\
Epochs & 100 \\
Evaluation Metrics & Micro-F$_1$, AUC, Accuracy \\
Baselines & Random GCN \\
Threshold Selection & F$_1$-optimized sweep (0.05–0.95) \\
\bottomrule
\end{tabular}
}
\caption{GCN architecture and training settings for future engagement prediction.}
\label{tab:engagement_gcn}
\end{table}

\begin{table}[h]
    \centering
    \renewcommand{\arraystretch}{1.1}
    \resizebox{0.48\textwidth}{!}{
    \begin{tabular}{llcc}
        \toprule
        \textbf{Time (days)} & \textbf{Method} & \textbf{F1} & \textbf{AUC} \\
        \midrule

        \multirow{8}{*}{t = 3}
        & Random GCN & 0.00 $\pm$ 0.00 & 0.00 $\pm$ 0.00 \\
        & Co-URL & 0.00 $\pm$ 0.00 & 0.00 $\pm$ 0.00 \\
        & Hashtag Sequence & 0.05 $\pm$ 0.01 & 0.42 $\pm$ 0.06 \\
        & Text Similarity & 0.01 $\pm$ 0.00 & 0.49 $\pm$ 0.04 \\
        & k-NN Embedding Graph & 0.01 $\pm$ 0.01 & 0.52 $\pm$ 0.05 \\
        & (Partial) Fused Graph & 0.03 $\pm$ 0.01 & 0.55 $\pm$ 0.05 \\
        & CANE & 0.29 $\pm$ 0.02 & 0.86 $\pm$ 0.03 \\
        & t-CANE & \textbf{0.34 $\pm$ 0.02} & \textbf{0.88 $\pm$ 0.03} \\
        \midrule

        \multirow{8}{*}{t = 5}
        & Random GCN & 0.00 $\pm$ 0.00 & 0.00 $\pm$ 0.00 \\
        & Co-URL & 0.00 $\pm$ 0.00 & 0.00 $\pm$ 0.00 \\
        & Hashtag Sequence & 0.08 $\pm$ 0.02 & 0.45 $\pm$ 0.07 \\
        & Text Similarity & 0.02 $\pm$ 0.01 & 0.53 $\pm$ 0.05 \\
        & k-NN Embedding Graph & 0.02 $\pm$ 0.01 & 0.57 $\pm$ 0.06 \\
        & (Partial) Fused Graph & 0.04 $\pm$ 0.01 & 0.60 $\pm$ 0.06 \\
        & CANE & 0.30 $\pm$ 0.02 & 0.87 $\pm$ 0.03 \\
        & t-CANE & \textbf{0.34 $\pm$ 0.02} & \textbf{0.88 $\pm$ 0.03} \\
        \midrule

        \multirow{8}{*}{t = 7}
        & Random GCN & 0.00 $\pm$ 0.00 & 0.56 $\pm$ 0.03 \\
        & Co-URL & 0.01 $\pm$ 0.00 & 0.43 $\pm$ 0.04 \\
        & Hashtag Sequence & 0.11 $\pm$ 0.04 & 0.48 $\pm$ 0.08 \\
        & Text Similarity & 0.02 $\pm$ 0.00 & 0.58 $\pm$ 0.05 \\
        & k-NN Embedding Graph & 0.02 $\pm$ 0.01 & 0.61 $\pm$ 0.06 \\
        & (Partial) Fused Graph & 0.05 $\pm$ 0.01 & 0.64 $\pm$ 0.06 \\
        & CANE & 0.30 $\pm$ 0.02 & 0.89 $\pm$ 0.02 \\
        & t-CANE & \textbf{0.35 $\pm$ 0.06} & \textbf{0.94 $\pm$ 0.02} \\
        \midrule

        \multirow{8}{*}{t = 14}
        & Random GCN & 0.00 $\pm$ 0.00 & 0.00 $\pm$ 0.00 \\
        & Co-URL & 0.00 $\pm$ 0.00 & 0.00 $\pm$ 0.00 \\
        & Hashtag Sequence & 0.06 $\pm$ 0.01 & 0.40 $\pm$ 0.06 \\
        & Text Similarity & 0.01 $\pm$ 0.00 & 0.47 $\pm$ 0.05 \\
        & k-NN Embedding Graph & 0.01 $\pm$ 0.01 & 0.51 $\pm$ 0.05 \\
        & (Partial) Fused Graph & 0.02 $\pm$ 0.01 & 0.53 $\pm$ 0.05 \\
        & CANE & 0.29 $\pm$ 0.02 & 0.86 $\pm$ 0.03 \\
        & t-CANE & \textbf{0.34 $\pm$ 0.02} & \textbf{0.88 $\pm$ 0.03} \\
        \bottomrule
    \end{tabular}
    }
    \caption{Performance of similarity networks for cross-platform narrative engagement prediction across time windows. Values reflect mean $\text{F}_1$ and AUC (± standard deviation) across validation folds. t-CANE consistently outperforms all baselines.}
    \label{tab:full-engagement-prediction}
\end{table}

\section{Appendix H: Fear Speech Detection and Examples}

\vspace{3pt}
\noindent
\textbf{Definition of Fear Speech.} Following~\citet{saha2023rise}, we define \textit{fear speech} as language that expresses apprehension or alarm about a group, portraying it as a threat to the in-group’s safety, values, or way of life. Unlike hate speech, which tends to be overtly derogatory or dehumanizing, fear speech is often framed as reasoned or cautionary discourse, designed to evoke anxiety or urgency. It may implicitly justify exclusionary or punitive responses while maintaining a rhetorically restrained tone~\cite{gerard2024fearloathingfrontlinedecoding}.

\vspace{3pt}
\noindent
\textbf{Examples From the Dataset.} \begin{quote}
\small
``Besides, they [immigrants] don’t believe our laws apply to them even when they come here. They’re already practicing p***philia in their enclaves.''
\end{quote}

\begin{quote}
\small
``Keep this one going, help the ones still blinded by communist regime. The media while Venezuelan gangs torture, kill citizens in 5 states! Biden/Kamala open borders—no one saw this on CNN, etc. Corrupt media. This was in Merit Street News!''
\end{quote}

\vspace{3pt}
\noindent
\textbf{Fear Speech Detection Model.} We follow the training setup and hyperparameter choices from~\citet{saha2023rise}, fine-tuning a RoBERTa-based classifier~\footnote{\url{https://huggingface.co/FacebookAI/xlm-roberta-large}} on their annotated dataset of over 22,000 social media posts labeled as hate, fear, or benign. The model is trained to identify group-directed alarmist rhetoric: \textit{e.g.,} language invoking cultural threat, physical danger, or national decline. \citet{saha2023rise} report a macro-F1 of 0.62 for their model on this task; our implementation yields comparable performance.

Each post in our dataset receives a continuous score between 0 and 1 indicating the likelihood of expressing fear speech. For our main analysis, we define fear speech as any post with a model score exceeding 0.75. This threshold, slightly more conservative than the default 0.5, prioritizes precision over recall to reduce false positives and ensure that identified posts reflect clearer expressions of fear-laden rhetoric. We verify that the same trend holds even under stricter criteria (up to thresholds of 0.95) and find that platform-origin differences persist. For example, when applying a stricter 0.9 cutoff, Truth Social-seeded narratives still exhibit a +0.23 log-odds ratio and a 23.5\% relative increase in fear-laden content compared to those originating from X ($p < 0.01$).

\section{Appendix I: Bridge Users}

\begin{figure}[ht]
    \centering
    \includegraphics[width=0.45\textwidth]{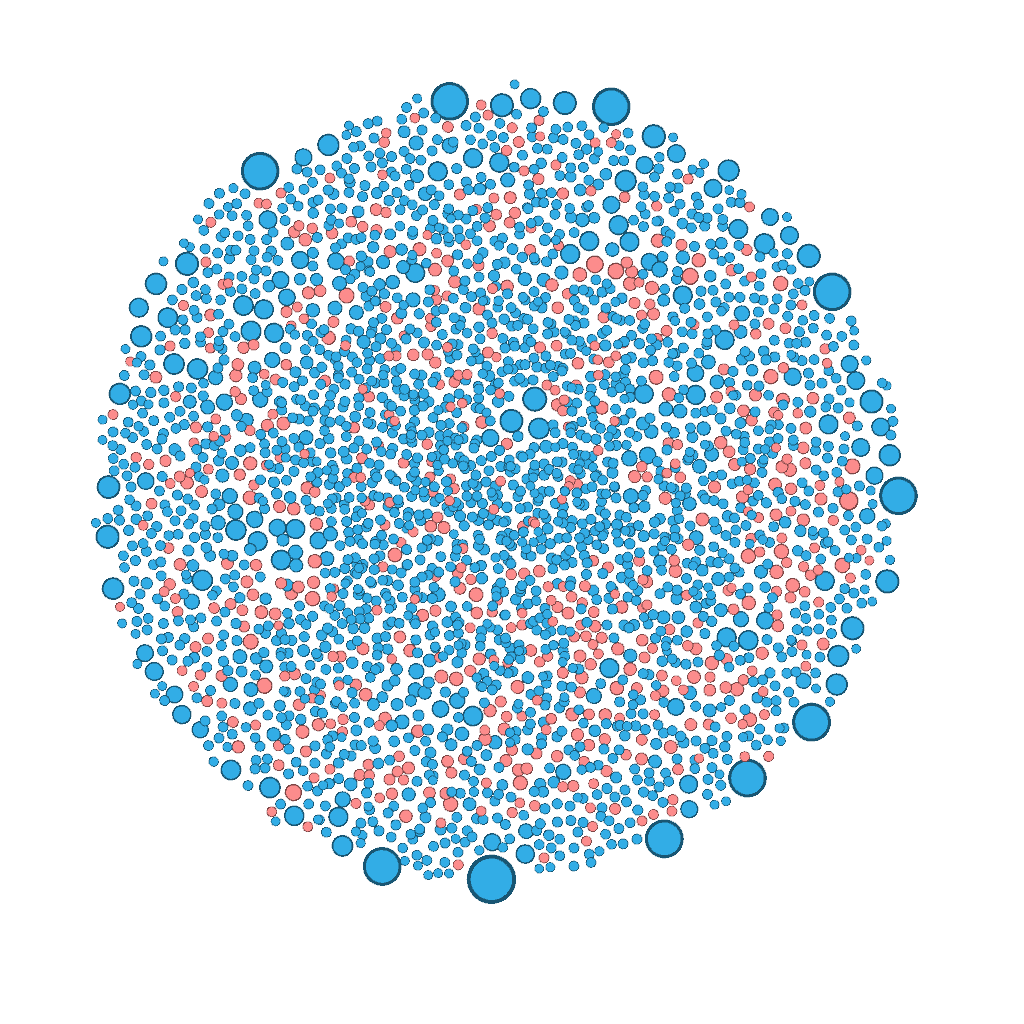}
    \caption{
        Visualization of the discourse network colored by platform (blue: X, red: Truth Social), with node size scaled by degree. This layout reveals a structurally embedded \textit{bridge zone}: a dense, mixed-platform region near the center where users from both platforms are highly interconnected. These users serve as key conduits for narrative migration across fragmented media environments. The concentration of red nodes within the blue-majority core illustrates cross-platform entanglement not evident in interaction-based graphs.
    }
    \label{fig:bridge-zone-visualization}
\end{figure}

\vspace{3pt}
\noindent
\textbf{Bridge Zone Visualization.} Figure~\ref{fig:bridge-zone-visualization} displays the bridge zone as an extracted subgraph from the full discourse network. Node positions are determined using a force-directed layout computed in Gephi~\footnote{\url{https://gephi.org/}} with the Force Atlas 2 algorithm. Each node is colored by platform (Truth Social in red, X in blue) and sized by degree. This view isolates a structurally integrated region where users from both platforms are densely connected, reflecting cross-platform alignment within the discourse network. We visualize node size using degree to highlight structural embeddedness: an indicator of how centrally users are positioned within the network topology.

\vspace{3pt}
\noindent
\textbf{Bridge User Activity Characteristics.}  
Table~\ref{tab:bridge_user_activity} reports median engagement metrics and within-platform percentile ranks for this subset. Given substantial differences in engagement norms and distributional skew between X and Truth Social, we use rank-based percentiles computed within each platform and then aggregate across them. We use rank-based percentiles to capture users' relative position in a way that is robust to outliers and platform-specific scale (because engagement distributions vary widely across platforms like X and Truth Social, and are often heavy-tailed, raw comparisons or z-scores can be misleading). Notably, bridge users are not uniformly high-volume or high-engagement accounts. Their engagement levels consistently hover near the platform medians, reinforcing that their influence stems not from visibility but from their structural positioning across fragmented discourse communities.

Notably, bridge users are not uniformly high-volume or high-engagement accounts. Their engagement levels (likes, replies, and reposts) consistently fall near the platform-level medians, with percentile ranks between the 47th and 58th percentiles. These results reinforce the idea that bridge users are not defined by activity level but by structural embeddedness: they serve as connectors across fragmented platform communities rather than high-visibility influencers.

\begin{table}[t]
    \centering
    \renewcommand{\arraystretch}{1.2}
    \resizebox{0.4\textwidth}{!}{ 
    \begin{tabular}{lcc}
        \toprule
        \textbf{Metric} & \textbf{Median} & \textbf{Bridge Percentile (Platform-Normalized)} \\
        \midrule
        Total Posts   & 14.00 & 58.49\% \\
        Reply Count   & 0.16 & 47.06\% \\
        Like Count    & 0.58 & 53.26\% \\
        Repost Count  & 0.02 & 55.19\% \\
        \bottomrule
    \end{tabular}
    }
    \caption{Median engagement metrics and corresponding platform-normalized percentile ranks for bridge users. Percentiles are calculated within each platform and then averaged.}
    \label{tab:bridge_user_activity}
\end{table}

\vspace{3pt}
\noindent
\textbf{Robustness Check on Low-Prominence Narratives.}
Although our matched schema likely already accounts for differences in engagement with viral or high-visibility content (confirmed via Mann--Whitney $U$ tests), we further restrict our analysis to test the robustness of this pattern. Specifically, we isolate a stricter subset of low-prominence narratives that, within the first 24 hours, attract no more than five users, span at most two communities, and appear on only one platform. In other words, narratives that exhibit minimal early signs of virality.

Even within this conservative subset, we observe large and statistically significant engagement differences. Clusters seeded early by bridge users receive substantially more likes ($p < 0.001$, rank-biserial $= 0.503$), replies ($p < 0.001$, $r = 0.432$), and reposts ($p < 0.001$, $r = 0.373$). These results confirm that bridge users are not simply interacting with content already trending toward virality. Rather, they are consistently associated with elevated engagement even when narratives begin with minimal visibility, suggesting that their influence operates independently of early popularity signals.

\vspace{3pt}
\noindent
\textbf{Robustness Checks on Early Seeding Thresholds.}  
To assess whether our findings regarding the impact of bridge users on engagement are sensitive to threshold selection, we replicate our main analysis across a range of early-seeding definitions. Specifically, we vary the user participation threshold from 5\% to 30\% (in increments of 5\%), while keeping the 24-hour temporal window fixed.

Across all thresholds, clusters seeded by bridge users consistently receive more likes, replies, and reposts than those seeded by matched non-bridge users. Effect sizes remain statistically significant and moderately large throughout, with rank-biserial correlations ranging from 0.27 to 0.60 for likes, 0.32 to 0.43 for replies, and 0.27 to 0.46 for reposts. These results demonstrate that our engagement-based findings are not an artifact of a particular threshold choice. Rather, they underscore the robustness of bridge users' outsized role in driving interaction, even under conservative definitions of early seeding.

\vspace{3pt}
\noindent
\textbf{Baseline Graph Evaluation.}
To assess whether other graph construction methods can uncover the same cross-platform alignment surfaced by our discourse-based approach, we construct several alternative user-user graphs: co-hashtag, co-URL, fused lexical similarity, and sentence-level embedding kNN graphs. Each graph is subjected to Louvain community detection, followed by platform entropy and narrative migration analysis.

Table~\ref{tab:efficiency-migratory} compares the proportion of users and posts identified as bridge users, along with the percentage of simple migrating narratives they introduced across platforms (we note that the trend remains the same when limited to significant migrating narratives):

\begin{table}[th]
\centering
\small
\resizebox{0.45\textwidth}{!}{ 
\begin{tabular}{lrrr}
\toprule
\textbf{Graph Type} & \textbf{\% Users} & \textbf{\% Posts} & \textbf{\% Narratives Introduced} \\
\midrule
Co-URL & 0.00\% & 0.00\% & 0.00\% \\
Hashtag Sequence & 0.04\% & 0.37\% & 4.21\% \\
Text Similarity & 1.42\% & 2.21\% & 15.70\% \\
kNN Embedding Graph & 1.73\% & 2.41\% & 25.90\% \\
Fused Graph & 2.68\% & 3.19\% & 27.00\% \\
\midrule
\textbf{Ours (Discourse)} & 0.33\% & 2.14\% & 67.72\% \\
\bottomrule
\end{tabular}
}
\caption{Comparison of bridge user detection and narrative introduction across graph construction methods.}
\label{tab:efficiency-migratory}
\end{table}

\vspace{3pt}
\noindent
Our method identifies 67.7\% of cross-platform narrative introductions using only 0.33\% of users: over 21$\times$ more efficient than the best-performing baseline (Fused Graph), which captures just 27.0\% using 2.68\% of users. Measured as the ratio of narratives introduced to users involved, our model achieves an efficiency score of \textbf{205}, far exceeding all other methods. Notably, the Hashtag Sequence graph performs poorly despite benefiting from feature leakage: hashtags were partially baked into the upstream data collection process by~\citet{balasubramanian2024public} and~\cite{shah2024unfiltered}. Both it and the co-URL graph fail to surface meaningful cross-platform bridges.

These findings underscore the distinctiveness of our bridge community and validate our clustering strategy, which surfaces discourse-level alignments rather than relying on shallow behavioral or lexical overlap. Our method reveals structurally coherent zones of narrative migration that remain invisible to traditional similarity-based approaches.

\vspace{3pt}
\noindent
\textbf{Bridge User Role in Cross-Platfrom Narrative Introduction.} To better understand the temporal role bridge users play in narrative diffusion, Table~\ref{tab:bridge_intro_thresholds} reports the proportion of clusters for which a bridge user appeared within the first $n$ posts on the receiving platform, across different cluster types. For example, among \textit{significant migrating clusters} (as identified by Pearson correlation), bridge users were the first to introduce the narrative 69.3\% of the time and were among the first three posters in 79.4\% of cases. These results suggest that bridge users are not simply eventual adopters, but consistent early initiators.

\begin{table}[h]
    \centering
    \renewcommand{\arraystretch}{1.15}
    \resizebox{0.48\textwidth}{!}{
    \begin{tabular}{l|ccccc}
        \toprule
        \textbf{Cluster Type} & \textbf{Top 1} & \textbf{Top 2} & \textbf{Top 3} & \textbf{Top 4} & \textbf{Top 5} \\
        \midrule
        Significant Migration & 69.33\% & 77.73\% & 79.41\% & 79.83\% & 80.67\% \\
        Simple Migration      & 67.72\% & 76.61\% & 78.29\% & 78.99\% & 79.45\% \\
        \bottomrule
    \end{tabular}
    }
    \caption{Percentage of cross-platform narrative clusters where a bridge user appeared within the first $n$ posts on the receiving platform.}
    \label{tab:bridge_intro_thresholds}
\end{table}

\end{document}